\RequirePackage{etex}
\documentclass{aa}

\usepackage{rotating}
\usepackage{graphicx}
\usepackage{float}
\usepackage{txfonts}
\usepackage{multirow}
\usepackage{appendix}
\usepackage{chngcntr}
\usepackage{textcomp}
\usepackage{silence}
\usepackage{subcaption}
\usepackage{pdflscape}
\usepackage{placeins}
\usepackage{natbib}
\bibpunct{(}{)}{;}{a}{}{,}
\usepackage[hidelinks]{hyperref}
\usepackage{orcidlink}

\begin{document}
\authorrunning{Hao et al.}
   \title{A cosmic filament catalog for COSMOS-Web Data Release 1 across $0.5 < z < 3.5$}

   \author{Qiaoyang Hao\inst{1,2}, Tao Wang\inst{1,2}\thanks{Corresponding author: taowang@nju.edu.cn}\orcidlink{0000-0002-2504-2421}, Weishan Zhu\inst{3,4}, Hong Guo\inst{5}, Peng Wang\inst{5}\orcidlink{0000-0003-2504-3835}, Tiancheng Yang\inst{1,2}, Luwenjia Zhou\inst{1,2}}
   \institute{School of Astronomy and Space Science, Nanjing University, Nanjing 210093, China
   \and Key Laboratory of Modern Astronomy and Astrophysics (Nanjing University), Ministry of Education, Nanjing 210093, China
   \and School of Physics and Astronomy, Sun Yat-Sen University, Zhuhai campus, No. 2, Daxue Road Zhuhai, Guangdong, 519082, China
   \and CSST Science Center for the Guangdong-Hong Kong-Macau Greater Bay Area, Daxue Road 2, 519082, Zhuhai, China
   \and Shanghai Astronomical Observatory, Chinese Academy of Sciences, Nandan Road 80, Shanghai 200030, China.}

   \date{Received date / Accepted date}
   \abstract
   {Cosmic filaments represent the environments in which most galaxies form and evolve. However, the spatial distribution of filaments and the link between galaxy properties and their positions relative to filaments remains unclear, particularly at high redshifts.}
   {In this work, using the deep-field catalog COSMOS-Web Data Release 1, we construct a filament catalog spanning $0.5 < z < 3.5$ that can be directly applied to studies of galaxy evolution and the galaxy--large-scale-structure connection at the peak epoch of galaxy and black hole formation.}
   {We developed a new procedure based on the DisPerSE method that fully incorporates the probability distribution functions (PDFs) of galaxy photometric redshifts by generating multiple realizations for each tomographic 2D slice. We then stacked these realizations to construct ``filament probability maps'' ($w_{\rm fil}$ maps), from which filaments were subsequently extracted. The filament extraction parameters were optimized using mock catalogs generated from TNG simulations by minimizing the discrepancies between 3D and projected 2D filaments.}
   {Applying this procedure to the COSMOS-Web data, we detected 12,523 filaments (comprising 136,210 segments) across $0.5 < z < 3.5$ and released a filament catalog, a companion segment catalog, and filament probability maps for each slice. Validation against astrophysical properties shows a strong agreement between the extracted filaments and those in the simulations, supporting the robustness of the filament extraction.}
   {These data products provide a robust foundation for investigating galaxy--environment coevolution at cosmic noon and beyond.}

   \maketitle

\section{Introduction} \label{sec:intro}

Understanding the distribution of galaxies and matter in space and time is a fundamental goal of extragalactic astronomy \cite{Dalmasso2024}. In the concordance $\Lambda$ cold dark matter ($\Lambda$CDM) paradigm, the large-scale structure (LSS) of the Universe emerges from the gravitational amplification of tiny primordial density fluctuations \citep{Peebles1967}: small overdensities grow under gravity, collapse hierarchically, and assemble into the cosmic web \citep{Bond1996,van_de_Weygaert2009}, comprising four components---nodes (clusters), filaments, sheets (walls), and voids \citep{Hayashi2019,Malavasi2020,Pasha2023,Porter2023,Peper2023}. Among these, filaments are particularly important despite their limited volume-filling fraction. They account for about 40\%--50\% of the baryonic matter in the Universe at $z=0$ \citep{Aragon2010,Cautun2014,Zhu2017}, form the backbone that links dense nodes, trace a large fraction of the tidal gravitational field in the Universe \citep{Kugel2024}, and thereby exert a dominant influence on galaxy growth and environmental quenching \citep{Yu2025}. At high redshifts, however, observational constraints on filaments remain scarce, motivating the need for robust high-$z$ filament catalogs.

During cosmic noon ($z \sim 2$), star formation and active galactic nucleus (AGN) activity peak \citep{Madau2014,Feldmann2017,Schreiber2020,Juodzbalis2025}, while interactions including mergers, gas accretion, and environmental quenching strongly drive galaxy evolution \citep{Morishita2014,Somerville2015,Bluck2022,Matteo2023,Park2024,Singh2025}.

At the same time, the LSS is still rapidly assembling itself \citep{Ata2021,Wang2024}, yet observational constraints on the cosmic web at these epochs remain limited. Because galaxy evolution and LSS are inseparable at these redshifts, studying high-redshift filaments is essential: they link LSS assembly to the processes that fuel star formation and sustain AGN activity, provide a test of how filaments supply cold gas and influence galaxy interactions and reveal how large-scale environments shape galaxy properties.

Multiple galaxy surveys have provided extensive evidence of filamentary structures in the local Universe and at low redshifts ($z \lesssim 0.5$), including SDSS \citep{York2000,Tempel2014}, 2dFGRS \citep{Cole2005}, and GAMA \citep{Driver2011}. At higher redshifts ($z \gtrsim 1$), filamentary structures have been traced using H$\alpha$ emitters \citep{Darvish2014}, Lyman alpha emitters (LAEs) \citep{Hayashino2004,Yamada2012} and Lyman Break Galaxies (LBGs) \citep{Hayashino2019}, but such tracers are largely confined to proto-cluster environments. Systematic studies of large-scale filaments near or before cosmic noon, however, remain largely limited to numerical simulations \citep{Zhu2021,Jhee2022,Lu2024}. Few surveys combine wide-area coverage, high-redshift depth, and precise 3D galaxy positions, leaving robust filament identifications at these epochs scarce. This scarcity limits direct tests of how the cosmic web during cosmic noon fuels star formation and sustains AGN activity, and of how it shapes galaxy evolution.

Nevertheless, considerable effort has been devoted to studying filaments in 2D tomographic slices \citep{Laigle2018,Duque2022,d'Assignies2025}, which extends such analyses to higher redshifts but entails substantial $z_{\rm phot}$ uncertainties and projection effects. These uncertainties can be mitigated by combining additional spectroscopic redshift ($z_{\rm spec}$) measurements with a probabilistic slice assignment for each galaxy \citep{Laigle2018,Laigle2025,EuclidCollab2025}. Although the projection from three to two dimensions inevitably introduces biases, morphological comparisons show that 2D filaments broadly trace the underlying 3D network \citep{Laigle2018}. Furthermore, the extent of these biases can be quantified by systematically comparing 3D and 2D filamentary networks \citep{EuclidCollab2025}. To date, however, no practical guideline has been established for selecting the parameters used to construct filament catalogs that can be directly applied to studies of the galaxy--large-scale-structure connection, particularly at high redshifts.

In this work, using the deep-field survey catalog COSMOS2025 (COSMOS-Web DR1; \citealt{Shuntov2025}), we present a new method that mitigates photometric-redshift uncertainties and projection biases in filament extraction.
The method has three main components: constructing a ``filament probability map'' ($w_{\rm fil}$ map) by stacking filaments extracted from multiple PDF($z$) realizations per tomographic slice; calibrating optimal extraction parameters through mock experiments with TNG300-1 simulations; and compiling a filament catalog that fully exploits the PDF($z$) while maximizing consistency with the 3D structures.
We extend the analysis across $0.5 < z < 3.5$ and provide a robust filament catalog and accompanying calibrated filament-probability maps directly applicable to studies of LSS and galaxy evolution.
In addition, we determine optimal choices of persistence thresholds for filament extraction, offering a quantitative reference for future studies, especially at high redshifts.

This paper is structured as follows. Section~\ref{sec:data} describes the observational and simulation data sets used in this work. Section~\ref{sec:methods} presents the filament extraction methodology, including the construction of tomographic slices and $w_{\rm fil}$ maps, the generation of mock catalogs, and the mock-driven optimization of persistence thresholds. Section~\ref{sec:res} presents the filament extraction results for both the mock catalog and the COSMOS-Web data, and validates the filament networks against their astrophysical properties. Finally, Section~\ref{sec:conclusions} summarizes our main results, discusses the caveats of the filament catalog, and compares our approach with conventional filament-extraction methods.

\section{Data} \label{sec:data}

\begin{table*}[htbp]
    \centering
    \caption{Summary of filament catalogs used in this work.}
    \label{tab:catalogs}
    {\renewcommand{\arraystretch}{1.3}
    \begin{tabular}{lllll}
    \hline
    Symbol & Catalog & Tracer & Dimensionality & Role \\
    \hline
    $\mathcal{F}_{\rm 3D,ref}$ & 3D TNG300-1 reference & Galaxies & 3D & Ground truth for threshold calibration \\
    $\mathcal{F}_{\rm 2D,ref}$ & 2D TNG300-1 reference & Galaxies & 2D & 2D-projected ground truth for threshold calibration \\
    $\mathcal{F}_{\rm 2D,mock}$ & 2D TNG300-1 ($w_{\rm fil}$) & Galaxies & 2D & Mock reconstruction and validation \\
    $\mathcal{F}_{\rm 2D,obs}$ & 2D COSMOS-Web ($w_{\rm fil}$) & Galaxies & 2D & Final observed filament catalog \\
    $\mathcal{F}_{\rm single}$ & Conventional single-$z$ & Galaxies & 2D & Baseline comparison (Sect.~\ref{sec:conclusions}) \\
    \hline
    \end{tabular}}
\end{table*}

\begin{figure*}[htbp]
    \centering
    \includegraphics[width=0.75\textwidth]{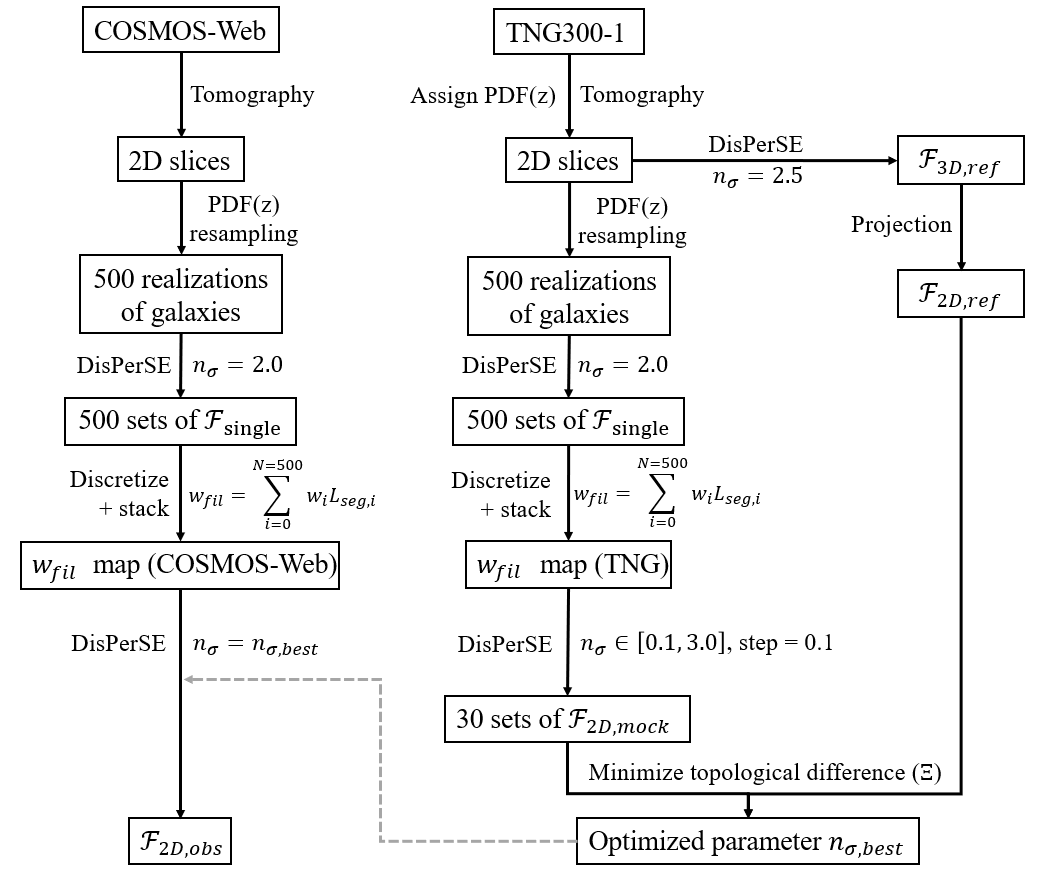}
    \caption{Overview of the COSMOS-Web filament catalog construction pipeline. The flow proceeds from the input data (COSMOS-Web DR1 and TNG300-1; left) through tomographic slicing and mock light-cone construction, $w_{\rm fil}$ map stacking, persistence threshold calibration via mock data, and DisPerSE re-extraction, to the final optimal filaments ($\mathcal{F}_{\rm 2D,obs}$).}
    \label{fig:flowchart}
\end{figure*}

\subsection{Observational data}\label{sec:obs_data}

The COSMOS-Web Survey \citep{Casey2023} is a 255-hour JWST Cycle 1 program covering 0.54 deg$^2$ with four NIRCam filters \citep{Rieke2023}, achieving a 5$\sigma$ point-source depth of 27.5--28.2 mag.
The COSMOS field \citep{Scoville2007,Capak07} benefits from a rich multiwavelength legacy spanning X-ray to radio. Optical and near-infrared coverage includes CFHT optical imaging \citep{Sawicki19}, high-resolution ACS-HST F814W imaging \citep{Koekemoer2007}, $grizy$-band imaging from Hyper Suprime-Cam (HSC), 13 intermediate and narrow bands from Subaru Suprime-Cam (SC) \citep{Taniguchi2015}, and the UltraVISTA survey \citep{McCracken2012,Moneti2023}, which provides $YJHK_{\mathrm{s}}$-band NIR imaging.

The COSMOS2025 catalog (COSMOS-Web DR1; \citealt{Shuntov2025}) is the latest version of the COSMOS catalog, covering the field with JWST NIRCam observations.
It contains 784,016 sources extracted from a PSF-homogenized $\chi^2$ detection image that combines all NIRCam filters, using the SourceXtractor++ model-fitting engine.
Photometric redshifts ($z_{\rm phot}$) for these sources are measured with the template-fitting code LePHARE \citep{Arnouts2002,Ilbert2006}. To exploit the diversity of multiband observations and the enhanced sensitivity of NIRCam, the galaxy template library has been expanded using BC03 models \citep{Bruzual2003}.
The resulting $z_{\rm phot}$ quality surpasses previous COSMOS catalogs, reaching a precision of $\sim$0.01 for $m_{\rm F444W} < 24$ with a 2\% catastrophic failure rate.
Even for the faintest galaxies at the highest redshifts ($26 < m_{\rm F444W} < 28$, $z \sim 3.5$)---the regime of interest for this work---the precision remains better than 0.03 with $\sim$10\% failure.

The COSMOS Spectroscopic Redshift ($z_{\rm spec}$) Compilation \citep{Khostovan2025} covers the 2 deg$^2$ COSMOS legacy field and comprises 165,312 redshifts for 97,929 unique objects from 108 observing programs, extending to $z \sim 8$ with median stellar masses of $\sim 10^9$--$10^{10}\,M_\odot$. Each source has a quality flag ($Q_f$) representing its confidence level (CL); we used only $z_{\rm spec}$ measurements with CL > 85\% ($Q_f = 3, 4,$ or $9$) for cross-matching with COSMOS-Web DR1, and assigned $z_{\rm spec}$ to the galaxy when available.

The COSMOS-Web deep galaxy group catalog \citep{Toni2025} contains 1,678 groups up to $z \sim 3.7$, with a purity of approximately 77\%. It reaches nearly two magnitudes deeper in member detection than the previous AMICO-based COSMOS group catalog \citep{Bellagamba2018}, benefiting from JWST's deep imaging, high spatial resolution, and high-quality photometry. In this work, we selected all groups within $0.5 < z < 3.6$ with $\texttt{DETECTION\_FLAG} < 20$, following the authors' recommendation for robust detections. The $M_{200c}$ and $R_{200c}$ estimates for these groups are derived in Section~\ref{sec:methods} and used for the connectivity--mass analysis in Section~\ref{sec:res}.

\subsection{Numerical simulation}\label{sec:sim}

To verify the completeness and purity of the extracted filament skeletons, we constructed a mock catalog from the IllustrisTNG simulations.
IllustrisTNG \citep{Pillepich2018} is a suite of cosmological hydrodynamical simulations featuring an updated galaxy formation model. We adopted TNG300-1, the largest volume in the suite $(205,000\,\mathrm{ckpc}/h)^3 \approx (302.6\,\mathrm{cMpc})^3$, with gas and dark-matter mass resolutions of $m_{\rm gas}=1.1\times10^7\,M_{\odot}$ and $m_{\rm DM}=5.9\times10^7\,M_{\odot}$, respectively. This volume is chosen to match the comoving extent of the observational tomographic slices (thickness $\sim$150 cMpc; Section~\ref{sec:tomo_slice}).
Stellar masses are taken as the total mass of stellar particles (PartType4) in each subhalo. We adopted cosmological parameters from Planck Collaboration \citep{Planck_Collaboration2016}, which included $\Omega_m=0.3089, \Omega_{\lambda}=0.6911,\Omega_b=0.0486$, and h=0.6774.

\section{Methods} \label{sec:methods}

This work uses five types of filament catalogs, summarized in Table~\ref{tab:catalogs}. The 3D TNG300-1 reference catalog, $\mathcal{F}_{\rm 3D,ref}$, provides the ground truth for threshold calibration, and $\mathcal{F}_{\rm 2D,ref}$ is its projected 2D counterpart. The 2D TNG300-1 mock catalog, $\mathcal{F}_{\rm 2D,mock}$, is produced with the same pipeline as the observations and is used for method validation. The 2D COSMOS-Web catalog, $\mathcal{F}_{\rm 2D,obs}$, is the final observed filament catalog. Finally, the conventional single-$z$ catalog, $\mathcal{F}_{\rm single}$, serves as a baseline comparison (Section~\ref{sec:comp_single}).

Combining the filament catalogs introduced above,
we constructed the final $\mathcal{F}_{\rm 2D,obs}$ through the
  pipeline illustrated in Fig.~\ref{fig:flowchart}.
  Starting from the COSMOS-Web observational data (Sect.~\ref{sec:obs_data}), we
  first divided the redshift range $0.5 < z < 3.5$ into 67 tomographic 2D slices
  (Sect.~\ref{sec:tomo_slice}) and then drew 500 realizations from each slice via
  PDF($z$) resampling (Sect.~\ref{sec:fila_extract}).
  The DisPerSE algorithm (Sect.~\ref{sec:disparse}) was applied to each realization
  at $n_\sigma = 2.0$ to extract filaments, yielding 500 sets of
  $\mathcal{F}_{\rm single}$; these were discretized and stacked to form the
  observed $w_{\rm fil}$ map (Sect.~\ref{sec:fila_extract}).

To re-extract filaments from this $w_{\rm fil}$ map with DisPerSE, we required
  an appropriate persistence threshold $n_{\rm \sigma,best}$---one that yields filament topology
  consistent with the 3D filaments projected to 2D.
  For calibration, we turned to the TNG300-1 mock data (Sect.~\ref{sec:mock_catalog}).
  Mock galaxies were assigned PDF($z$) distributions statistically consistent with
  the observed $\Delta z$ behavior (Sect.~\ref{sec:optimize}), and an identical
  pipeline---slicing, resampling, $n_\sigma = 2.0$ extraction, discretization,
  and stacking---was applied to produce the mock $w_{\rm fil}$ map.

Meanwhile, the known 3D galaxy positions in the simulation allow us to extract $\mathcal{F}_{\rm 3D,ref}$ directly from the 3D galaxy distribution and project it to 2D as $\mathcal{F}_{\rm 2D,ref}$, which serves as the ground truth for topology comparison. From the mock $w_{\rm fil}$ map, we extracted 30 sets of $\mathcal{F}_{\rm 2D,mock}$ at $n_\sigma = 0.1, 0.2, \ldots, 3.0$, and compared each against $\mathcal{F}_{\rm 2D,ref}$ using the similarity metric $\Xi$ (Sect.~\ref{sec:optimize}). The $n_\sigma$ that minimizes $\Xi$ was identified as $n_{\sigma,\rm best}$. Finally, applying $n_{\sigma,\rm best}$ to the observed $w_{\rm fil}$ map yielded $\mathcal{F}_{\rm 2D,obs}$, the final observed filament catalog whose properties and astrophysical validation are presented in Sect.~\ref{sec:res} and discussed in Sect.~\ref{sec:conclusions}.

\subsection{Persistent skeleton algorithm (DisPerSE)}\label{sec:disparse}

DisPerSE \citep{Sousbie2011,Sousbie2011II} extracts filamentary structures as ridges of the density field, and has been shown to outperform other filament-finding methods in comparative studies \citep{Libeskind2018}. The density field is estimated via the Delaunay tessellation field estimator (DTFE, \cite{Schaap2000}), applied to the discrete galaxy distribution. Filaments are defined as the lines connecting topologically robust saddle--peak critical point pairs (persistence pairs); each such line is termed a segment. The topological significance of each segment is quantified by the persistence of its corresponding point pair, and the persistence value serves as a robustness measure that enables noise-resistant extraction through low-persistence filtering.

DisPerSE provides two parameters for setting the persistence threshold. The \texttt{-nsig} parameter ($n_\sigma$) is designed for discrete density fields estimated via DTFE; it defines the threshold in units of $\sigma$ relative to the noise level, with the noise probability given by $p = \frac{1}{2}\mathrm{erfc}(n_\sigma/\sqrt{2})$, where $\mathrm{erfc}$ is the complementary error function, describing the probability that a Gaussian noise fluctuation exceeds $n_\sigma$ standard deviations. The \texttt{-cut} parameter applies an absolute persistence threshold for continuous density fields on regular grids.

In this study, the 3D reference filament skeleton was extracted using $n_\sigma$=2.5, corresponding to a noise probability of $\sim$0.6\%. For the 2D reconstructions, we retained the $n_\sigma$ convention by mapping it to an equivalent \texttt{-cut} value via the empirical persistence distribution of random-background fields (Section~\ref{sec:optimize}). Unless otherwise specified, all filaments were smoothed by applying a moving average over ten segments without modifying critical point positions or segment counts.

\subsection{Tomographic slices generation}\label{sec:tomo_slice}

\begin{figure*}[htbp]
    \includegraphics[width=1\textwidth]{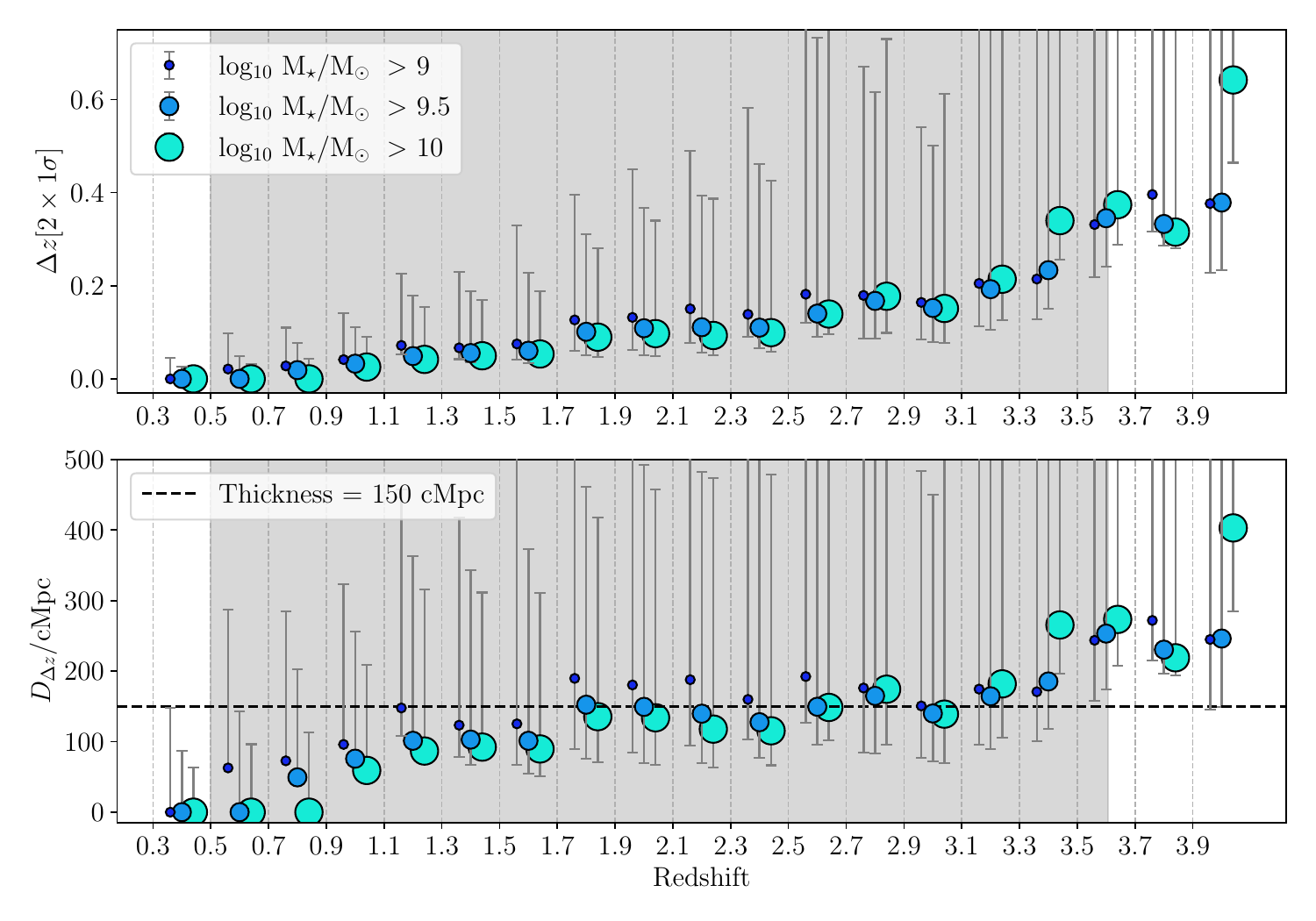}
    \caption{$z_{\rm phot}$ error and corresponding line-of-sight positional uncertainties ($D_{\rm \Delta z}$) of galaxies above different mass thresholds with respect to their redshifts. Median values and error bars are calculated for each redshift bin marked with vertical dashed lines.}
    \label{fig:z_check}
\end{figure*}

For each galaxy, the redshift probability distribution function (PDF($z$)) defines the photometric redshift uncertainty: $\Delta z$ is taken as the half-width of the 68\% credibility interval of PDF($z$). The median $\Delta z$ of galaxies with $M_{\star} > 10^{10}\,M_\odot$ is $\sim$0.02 at $z\sim 1.5$ (Figure~\ref{fig:z_check}), corresponding to $\sim$50 cMpc in comoving space---making a direct 3D reconstruction of the cosmic web infeasible. We therefore extracted filaments in 2D tomographic slices whose thickness is set by the typical $z_{\rm phot}$ error of the full galaxy sample.

We selected galaxies by removing non-galaxy objects (\texttt{type}=0 or \texttt{warn\_flag}=0), objects fainter than $m_{\rm 444w} = 30$, and those falling within the HSC bright-star mask (\texttt{flag\_star\_hsc}=0; \citealt{Coupon2018}). The remaining galaxies were cross-matched with the COSMOS spectroscopic redshift compilation \citep{Khostovan2025}; for matched galaxies we set $\Delta z = 0$, since their spectroscopic redshifts are effectively exact compared to the photometric uncertainties.

Figure~\ref{fig:z_check} shows the line-of-sight comoving uncertainty $D_{\rm \Delta z}$ corresponding to $\Delta z$ as a function of redshift for several stellar-mass thresholds. For $M_{\star} > 10^{9.5}\,M_\odot$, the median $D_{\rm \Delta z}$ rises from nearly zero at $z\sim 0.5$ to $\sim$200 cMpc at $z\sim 3.5$, with the majority of values remaining below 150 cMpc (horizontal dashed line). A slice thickness of 150 cMpc thus captures approximately 68\% of the PDF($z$) probability for a typical galaxy---corresponding to the $\pm 1\sigma$ confidence interval---and we adopted this as the slice thickness. We divided the comoving volume between $z=0.5$ and $z=3.5$ into 67 tomographic slices of thickness 150 cMpc with an overlap of 75 cMpc between adjacent slices. For detailed information about these slices, please see Appendix~\ref{app:detail}.

For statistical analyses, the 67 slices are grouped into four redshift bins: $0.50<z<0.82$, $0.82<z<1.21$, $1.21<z<1.91$, and $1.91<z<3.52$. These boundaries are chosen so that each bin spans approximately 1.8 Gyr of cosmic time, ensuring that galaxies in different bins are observed over comparable evolutionary intervals. This yields 7, 6, 8, and 12 even slices per bin, respectively, with the increasing number of slices at higher redshift reflecting the growing comoving volume per unit redshift. All redshift-dependent results in this work, including filament extraction thresholds, galaxy stellar mass functions, mass gradients, and connectivity--mass relations, are reported within these four bins. With the stellar mass cut $M_{\star} > 10^{9.5}\,M_\odot$, the redshift range $0.5 < z < 3.5$, and the selection criteria described above, the total sample comprises 37,298 galaxies, of which 6,678 ($\sim$17.9\%) have spectroscopic redshifts with a confidence level of above 85\%. Throughout this paper, galaxy coordinates are rotated clockwise around a center located at R.A. = 150.11974$^\circ$ and Decl. = 2.20845$^\circ$, with a rotation angle of $\theta$ = 19.993$^\circ$, for visual clarity in the illustrations.

\subsection{Filament extraction method}\label{sec:fila_extract}

\begin{figure*}[htbp]
    \includegraphics[width=1\textwidth]{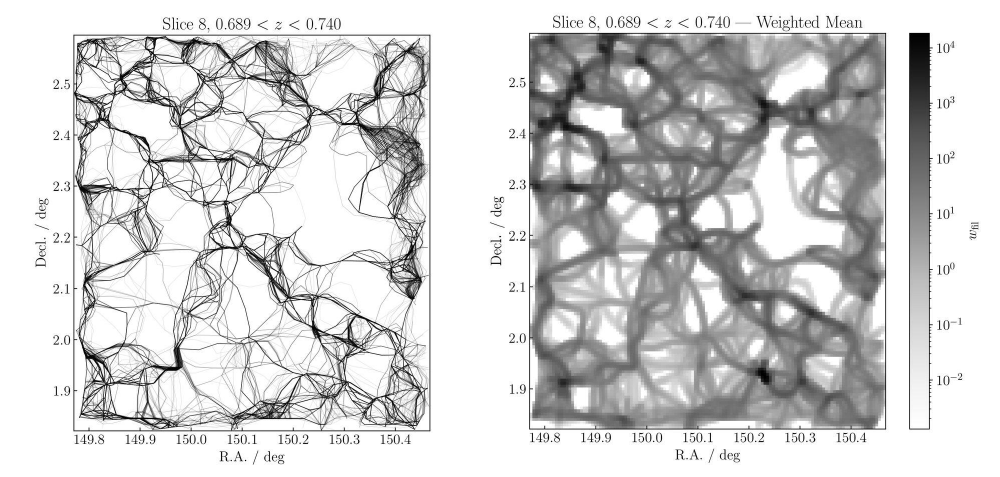}
    \caption{Stacked illustration of filaments extracted from all 500 realizations (left panel) and its corresponding $w_{\rm fil}$ map with a resolution of 250 ckpc (right panel) of the 8th slice in the observed sample.}
    \label{fig:wfil_map}
\end{figure*}

Tomographic slicing assigns each galaxy to a slice based on its best-estimate $z_{\rm phot}$, but the finite $z_{\rm phot}$ uncertainty means a galaxy's true redshift may lie in a neighboring slice. Moreover, galaxies with limited photometric coverage or degenerate redshift solutions can have PDF($z$)s that are broad, multi-peaked, or irregular, spanning several slices in the line-of-sight direction. The full shape of the PDF($z$) must therefore be taken into account alongside the tomographic slicing. Laigle et al. \citep{Laigle2025} addressed this by generating 100 redshift realizations per slice, drawing redshifts from each galaxy's PDF($z$), and analyzing the distribution of galaxy environment statistics across all realizations.

Building on this approach, we generated 500 redshift realizations per slice by drawing from each galaxy's PDF($z$) and extracted filaments from each realization using DisPerSE with $n_\sigma$=2.0. The resulting filament maps from all 500 realizations were stacked into a single composite map (Figure~\ref{fig:wfil_map}), yielding a filament probability map that is more robust to $z_{\rm phot}$ uncertainties than any single realization.

For the stacking procedure, we extended the sample beyond the 37,298 galaxies selected in Section~\ref{sec:tomo_slice} by also including galaxies whose PDF($z$) has nonzero support within $0.5 < z < 3.5$ even if their best-estimate $z_{\rm phot}$ lies outside this interval. The expanded sample comprises 41,051 galaxies.

To quantify the likelihood that a given region contains filaments, we discretized the stacked map onto a grid of 250 ckpc resolution, and assigned each cell a filament weight

\begin{equation}
    w_{\rm fil} = \sum_i^N L_{{\rm seg},i}\,w_i,
\end{equation}

where $N=500$ represents the number of realizations, $L_{{\rm seg},i}$ is the length of the filament segment in the $i$-th realization within the cell, and $w_i$ is the mean robustness\footnote{In DisPerSE, robustness is a dimensionless local measure of how contrasted a filament segment is with respect to its background density field. Higher values indicate more prominent structures.} of its two endpoints as reported by DisPerSE. We then redistributed the $w_{\rm fil}$ field via cloud-in-cell (CIC) interpolation to suppress grid-locking artifacts. The resulting $w_{\rm fil}$ map is shown in the right panel of Figure~\ref{fig:wfil_map}.

The $w_{\rm fil}$ map thus represents the likelihood of filament presence inside each $250^2$ (ckpc)$^2$ cell, incorporating the full PDF($z$) information of each galaxy. By re-extracting filaments from the $w_{\rm fil}$ maps at an optimized persistence threshold (Section~\ref{sec:optimize}), we obtained a final filament catalog that is expected to reduce the systematic offsets between 2D slice-based filaments and their projected 3D counterparts.

\subsection{Mock catalog generation}\label{sec:mock_catalog}

To determine the optimal persistence threshold for filament extraction from the $w_{\rm fil}$ maps, we calibrated against a mock experiment using the TNG300-1 simulation. We constructed a mock light cone that matches the comoving volume, slice geometry, and selection criteria of the observational data (Section~\ref{sec:tomo_slice}), then applied the same filament extraction pipeline to both the mock data and a 3D reference skeleton to identify the threshold that maximizes consistency between the two.

The construction of the mock catalog proceeded in three steps. First, we built a light cone from the TNG300-1 simulation box whose geometry and comoving volume were matched to the COSMOS-Web observational slices (Sect.~\ref{sec:tomo_slice}). Second, we populated the light cone with a mock galaxy sample and assigned photometric redshift uncertainties calibrated against the observed $\Delta z$--$z$--$m_{\rm Ks}$ relation. Third, we incorporated a group and cluster sample to enable the connectivity--mass validation. The three steps are detailed below, and the calibration of the persistence thresholds against the resulting mock data is described in Section~\ref{sec:optimize}.

\subsubsection{Mock light cone generation}

\begin{figure*}[htbp]
    \includegraphics[width=1\textwidth]{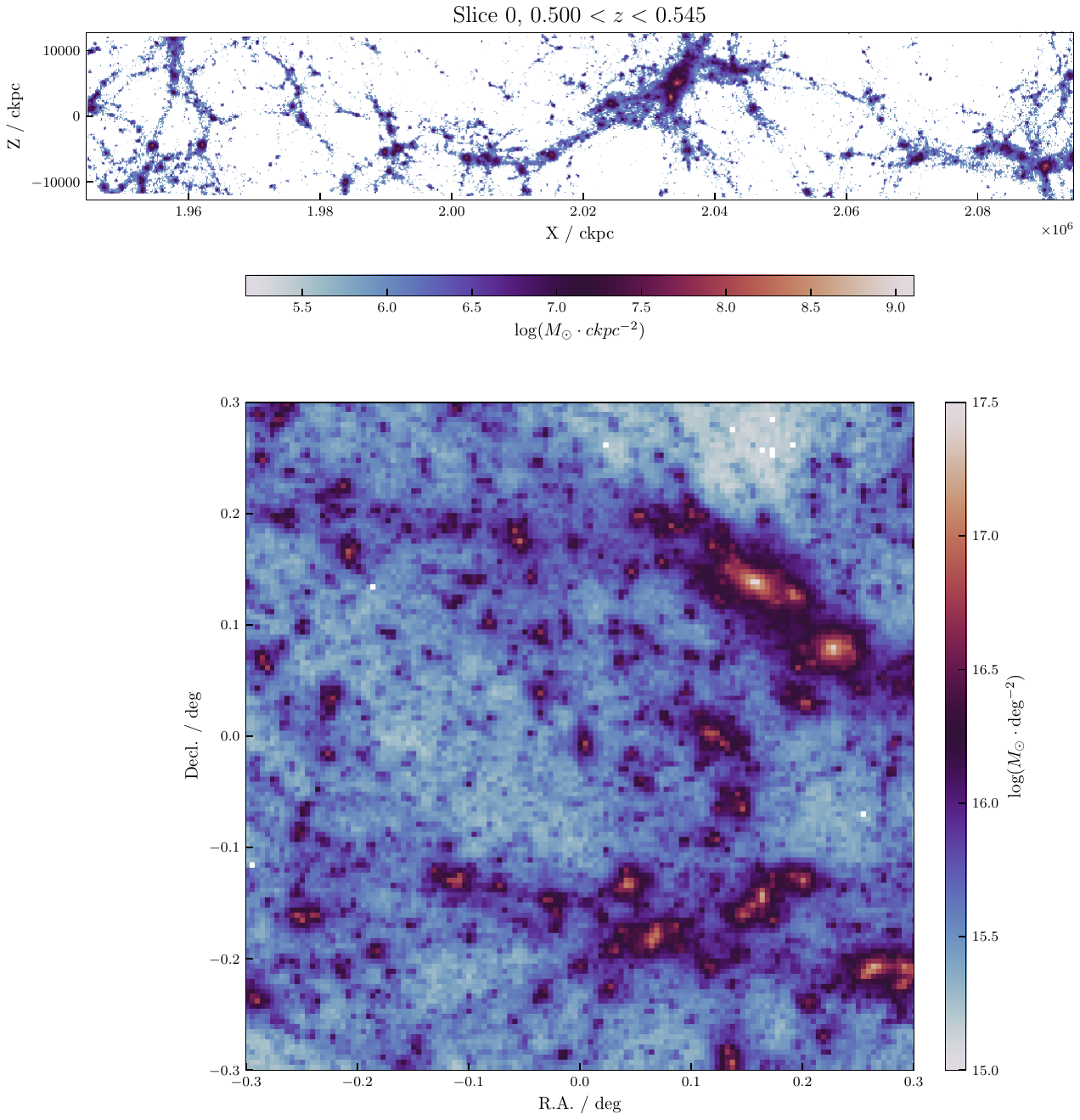}
    \caption{Side view of the dark matter distribution of the first mock light-cone slice extracted from the TNG300-1 simulation box in comoving space (upper panel) and the corresponding 2D projection from the front view (lower panel). Colors indicate the surface density of dark matter, in units of log($\mathrm {M_{\odot}\cdot ckpc^{-2}}$) (upper panel) and log($\mathrm{M_{\odot}\cdot deg^{-2}}$) (lower panel). For the upper panel, labels X and Z represent the Cartesian coordinates ($x_{ob}$,$z_{ob}$) that originate at the observer.}
    \label{fig:heatmap_dm}
\end{figure*}

The mock light cone was generated following the procedure of \citet{Bernyk2016}. To avoid artificial periodicity from simply tiling replicated copies of the simulation box, we applied random rotations, mirroring, and translations to the full box before extracting each sub-volume. Only even-numbered slices were used to ensure non-overlapping sub-boxes, and the resulting geometry was matched to the COSMOS-Web field of view. Boundary discontinuities between successive slices have amplitudes below 200 ckpc (see Figure D1 in \citet{Pichon2010}), smaller than the 250 ckpc resolution of the $w_{\rm fil}$ maps, so they do not significantly affect the filament extraction statistics or the optimal threshold determination. The full construction procedure is given in Appendix~\ref{app:lightcone}.

\subsubsection{Mock galaxy sample generation}

Galaxies in the mock catalog are represented by subhalos identified with SUBFIND and FoF \citep{Springel2005,Dolag2009}; stellar masses are defined in Section~\ref{sec:sim}. To match the tracer number density of the observed sample, we adopted a stellar mass threshold of $10^{9.7}\,M_{\odot}$ \citep{Laigle2018}---slightly higher than the observational limit ($10^{9.5}\,M_{\odot}$, Section~\ref{sec:tomo_slice}) to account for the incompleteness of the simulated galaxy population at the low-mass end. The resulting mock sample comprises 45,529 galaxies, closely matching the observed sample size before the HSC star mask was applied.

\begin{figure}
    \includegraphics[width=0.48\textwidth]{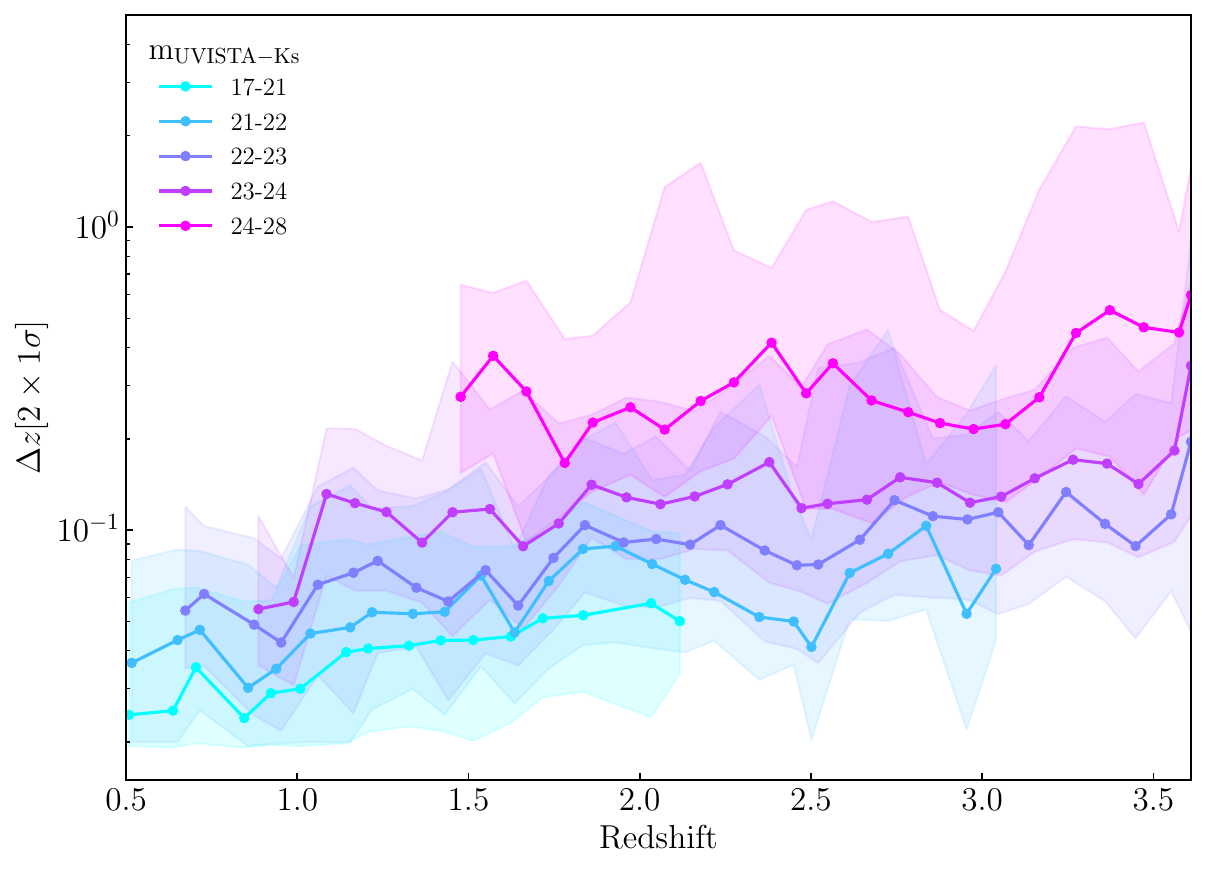}
    \caption{Variation in the median values of $z_{\rm phot}$ error $\Delta z$ with Ks-band magnitudes and redshifts in the observed sample. Shaded regions represent the uncertainties between the 16th and 84th percentiles.}
    \label{fig:zerr_obs}
\end{figure}

\begin{figure*}[htbp]
    \includegraphics[width=1\textwidth]{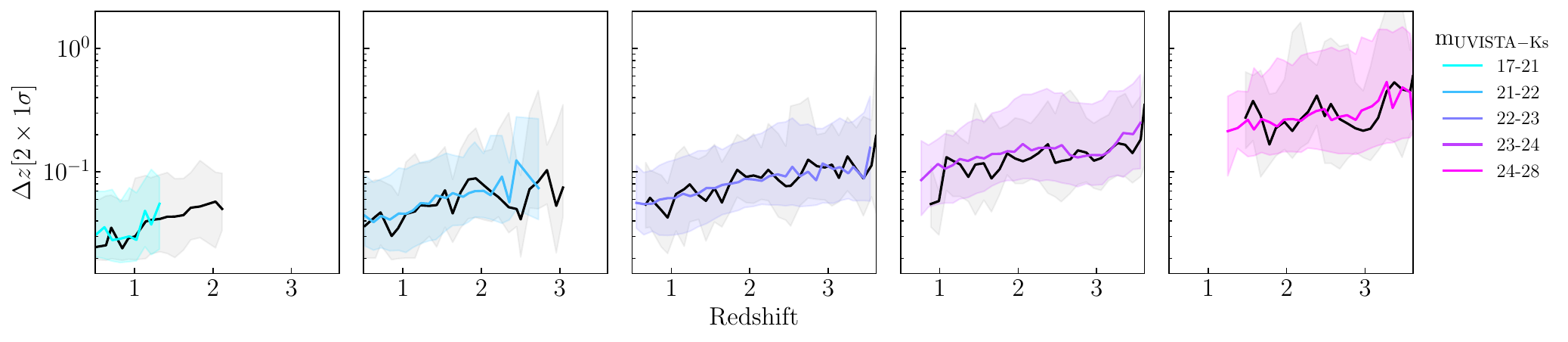}
    \caption{Validation of the reproduced $\Delta z$ in the mock galaxy sample. Panels from left to right correspond to increasing Ks-band magnitudes. In each panel, the solid black line and shaded region denote the median value and associated uncertainty of the observed sample, consistent with those presented in Figure~\ref{fig:zerr_obs}, while the colored line and shaded region denote the median value and associated uncertainty of the mock-reproduced sample.}
    \label{fig:zerr_check}
\end{figure*}

The cosmological redshift $z_0$ of each galaxy follows directly from its comoving distance $|\mathbf{r}_{\rm ob}|$. The simulated spectroscopic redshift $z_{\rm spec,sim}$ includes the contribution of peculiar velocities,

\begin{equation}
    z_{\rm spec,sim} = z_0 + \frac{\mathbf{v}_{\rm ob} \cdot \mathbf{r}_{\rm ob}}{|\mathbf{r}_{\rm ob}|}\,\frac{1 + z_0}{c},
\end{equation}

where $\mathbf{v}_{\rm ob}$ is the physical peculiar velocity of the galaxy \citep{Ishikawa2024}. To emulate photometric redshift uncertainties, we assigned each mock galaxy a Gaussian PDF($z$) centered at $z_{\rm spec,sim}$. This Gaussian form is a simplification; the limitations of this choice are discussed in Section~\ref{sec:conclusions}.

To assign a photometric redshift uncertainty to each mock galaxy, we needed its Ks-band magnitude, which is the observable that best correlates with $\Delta z$ in the COSMOS-Web data (Figure~\ref{fig:zerr_obs}). We computed $m_{\rm Ks,sim}$ using the \texttt{GALAXEV} package \citep{Rodriguez-Gomez2019}, treating each stellar particle as a single stellar population with BC03 models \citep{Bruzual2003}, a Chabrier IMF \citep{Chabrier2003}, and the Charlot \u0026 Fall dust attenuation prescription \citep{Charlot2000}.

We binned the observed sample in redshift and $m_{\rm Ks}$ to map the distribution of $\Delta z$ (Figure~\ref{fig:zerr_obs}), and used a Radial Basis Function (RBF) interpolant to estimate the median $\Delta z$ and its 16th--84th percentile range for each mock galaxy given its $z_{\rm spec,sim}$ and $m_{\rm Ks,sim}$. For each galaxy, we randomly drew an adopted $\Delta z$ from the interval $[\Delta z_{16}, \Delta z_{84}]$, following a skewed distribution, and constructed a Gaussian PDF($z$) centered at $z_{\rm spec,sim}$ with width $\sigma = \Delta z$.

We verified that this procedure reproduced the observed $\Delta z$--$m_{\rm Ks}$--$z$ relation by applying it to the mock sample and comparing with the data (Figure~\ref{fig:zerr_check}). Finally, to mimic the observed $z_{\rm spec}$-complete fraction, we set $\Delta z = 0$ for the same fraction of bright galaxies as in the COSMOS-Web sample.

\subsubsection{Mock group and cluster sample generation}

To consider the effect of groups and clusters when extracting filaments and validate the filament--cluster connectivity analysis (Section~\ref{sec:res}), we incorporated a cluster sample into the mock catalog. Galaxy groups were identified with a standard friends-of-friends (FoF) algorithm with linking length $b=0.2$; we treated groups with $R_{200c} > 0.5$ cMpc as clusters. Every halo whose minimum-potential position ($\texttt{GroupPos}$) fell within the extracted sub-box was selected. We adopted a radius-based selection because $R_{200c}$ directly controls whether a cluster's spatial extent is resolved by the $w_{\rm fil}$ map (resolution 250 ckpc); the threshold is set at 500 ckpc ($2\times$ the grid resolution) to ensure that selected clusters are spatially distinct from filamentary structures.

\subsection{Parameter optimization for filament extraction}\label{sec:optimize}

The primary objective of the mock catalog is to calibrate the persistence thresholds used for filament extraction from the $w_{\rm fil}$ maps, so as to identify filaments with high reliability in the 2D observational slices. Persistence, as described in \cite{Sousbie2011}, quantifies the lifespan of a topological feature across a filtration, defined as the difference or ratio between the field values at the birth and death critical points of that feature; a larger persistence indicates a more robust, less noise-like structure.

DisPerSE expresses structural significance in two complementary modes, depending on the input data format. For density fields computed via the DTFE on discrete point distributions---such as the galaxy positions used to construct the 3D reference skeleton---the \texttt{-nsig} parameter characterizes the probability $p$ that a structure arises from random Poisson noise: $p = \frac{1}{2}\,\mathrm{erfc}(n_{\sigma}/\sqrt{2})$. For density fields defined on regular grids, such as the $w_{\rm fil}$ maps, DisPerSE uses the \texttt{-cut} parameter, which applies an absolute persistence threshold. To ensure consistent statistical meaning between the two modes, we mapped $n_{\sigma}$ to a \texttt{-cut} value via the empirical cumulative distribution function (ECDF) of persistence values measured from random-background realizations of the $w_{\rm fil}$ maps: for a target $n_{\sigma}$, the corresponding noise probability $p$ is computed, and the cut is set to $\mathrm{ECDF}^{-1}(1-p)$---the persistence value above which a fraction $p$ of random fluctuations would be incorrectly identified as signal.

We fixed the threshold of $\mathcal{F}_{\rm 3D,ref}$ extraction at $n_{\sigma} = 2.5$, corresponding to $p \approx 0.6\%$. This value is a standard choice for tracing filaments from discrete galaxy distributions \citep{Laigle2018,Wang2024,EuclidCollab2025}. The free parameter to be optimized is the predicted threshold $n_{\sigma,\rm pred}$, which controls the persistence level for re-extraction from the $w_{\rm fil}$ map after conversion to a \texttt{-cut} value via the ECDF mapping. We scanned $n_{\sigma,\rm pred} \in [0.1, 3.0]$ in steps of 0.1: for each candidate, we converted it to \texttt{-cut}, re-extracted $\mathcal{F}_{\rm 2D,mock}$, and evaluated the filament-similarity metric $\Xi$ \citep{EuclidCollab2025} against $\mathcal{F}_{\rm 3D,ref}$. The $\Xi$ metric jointly quantifies the distance and angular separations between nearest-neighbor segment pairs in the reference and predicted skeletons, computed in both 3D$\rightarrow$2D and 2D$\rightarrow$3D directions; the $n_{\sigma,\rm pred}$ that minimizes $\Xi$ is identified as $n_{\sigma,\rm pred,best}$. Unless otherwise stated, the optimization restricts the reference skeleton to its top 50\% in robustness; results without this restriction are presented in Appendix~\ref{app:zero}. Before calibrating the thresholds, we verified that the mock and observed $w_{\rm fil}$ maps yielded statistically consistent persistence distributions (Appendix~\ref{app:persis_compare}), ensuring that thresholds derived from the mock can be applied to the data. Full details of the scan setup, the $n_{\sigma}{\rightarrow}\texttt{cut}$ mapping, and the $\Xi$ metric are provided in Appendix~\ref{app:thresh_detail}.

\begin{figure*}[htbp]
    \includegraphics[width=1\textwidth]{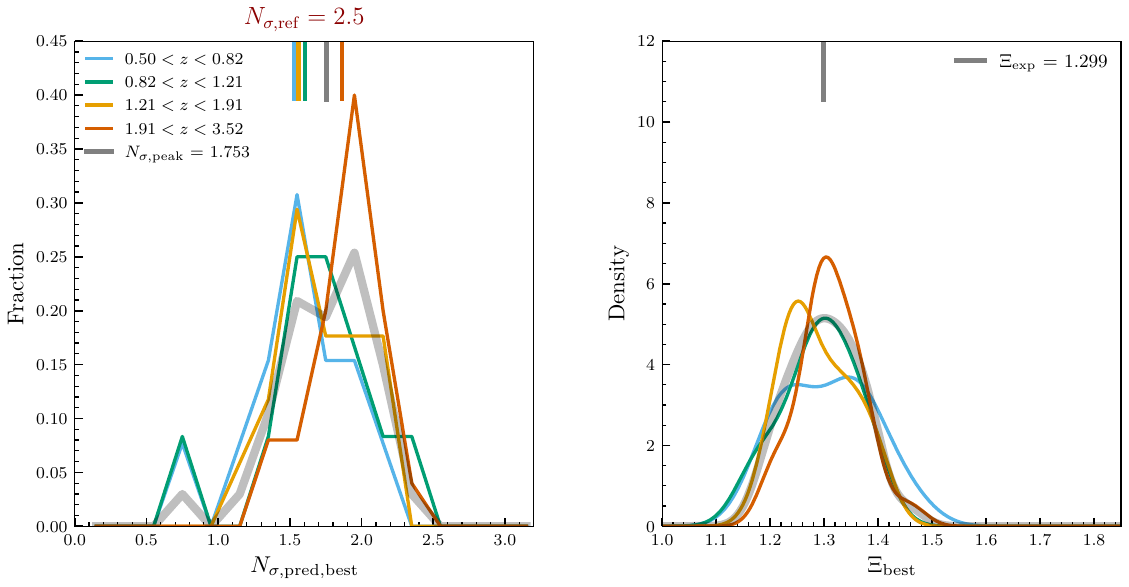}
    \caption{Distributions of $n_{\sigma,\rm pred,best}$ (left) and the corresponding density distributions of $\Xi_{\rm best}$ (right) obtained via kernel density estimation (KDE), for $n_{\sigma,\rm ref} = 2.5$, considering the 50\% most robust reference filaments. Different colors denote redshift bins. The peak values for each group are determined by KDE and indicated by short line segments at the top of each panel, for the full sample (gray line) and for individual redshift bins (colored lines).}
    \label{fig:pred_opt_r50_nomw}
\end{figure*}

\begin{figure*}[htbp]
    \includegraphics[width=1\textwidth]{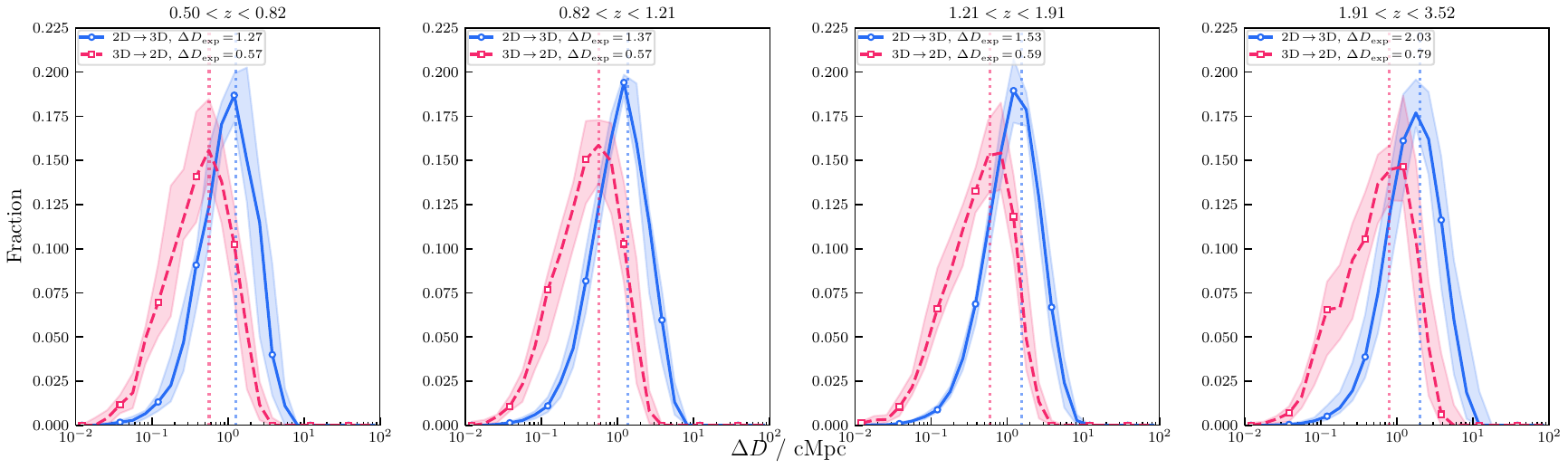}
    \caption{Distance separation distributions at the optimal $n_{\sigma,\rm pred}$ for each redshift bin, with $n_{\sigma,\rm ref} = 2.5$ and the 50\% most robust reference filaments. Different columns represent different redshift bins, with increasing redshift from left to right. Red (blue) lines refer to 3D $\rightarrow$ 2D (2D $\rightarrow$ 3D). The shaded region indicates the 16th--84th percentile range within each group. Expectation values are marked as short vertical lines above each panel, with values shown in the legend.}
    \label{fig:dist_dis}
\end{figure*}

\begin{figure*}[htbp]
    \includegraphics[width=1\textwidth]{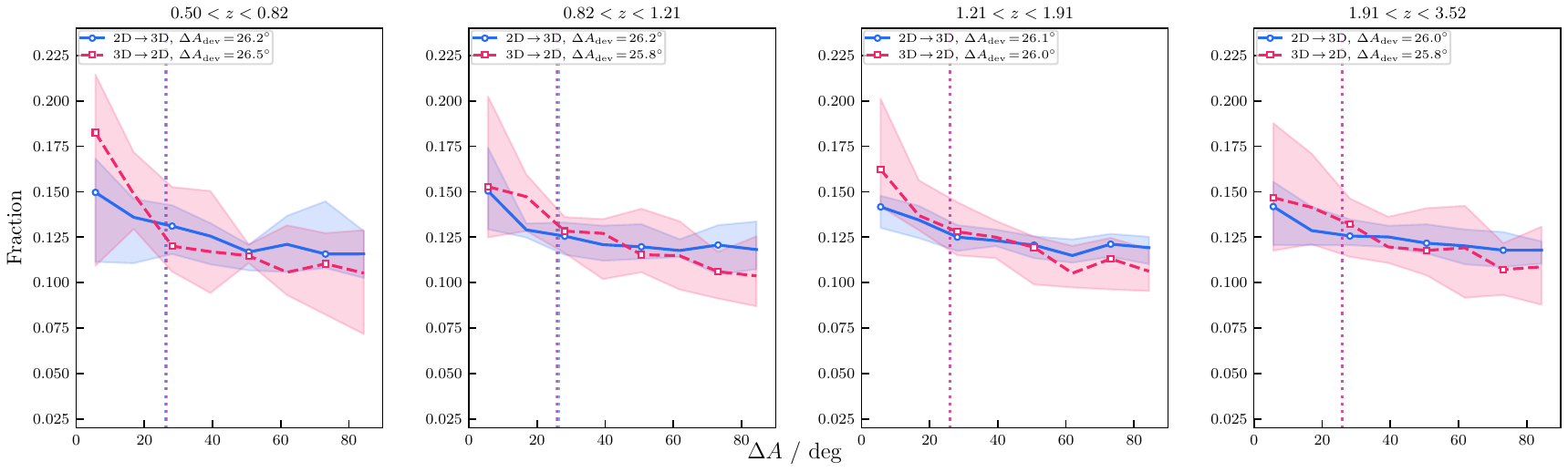}
    \caption{Similar to figure~\ref{fig:dist_dis} but showing angular separation distributions at the optimal $n_{\sigma,\rm pred}$ for each redshift bin.}
    \label{fig:ang_dis}
\end{figure*}

\begin{figure*}[htbp]
\centering
    \includegraphics[width=1\textwidth]{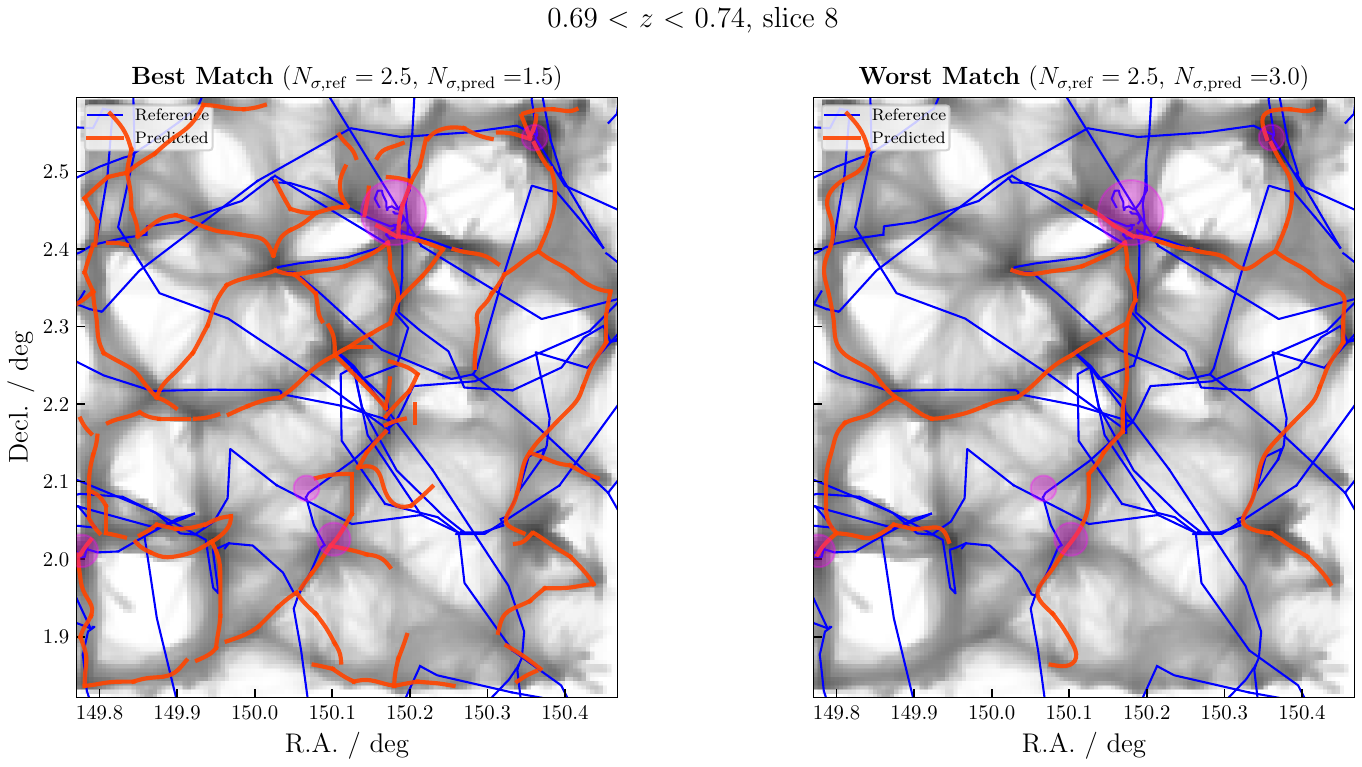}
    \caption{Illustration of filament extraction results for slice 8 (0.69 < $z$ < 0.74) of the TNG300-1 mock catalog at the best-match $n_{\sigma,\rm pred}$ (left) and the worst-match $n_{\sigma,\rm pred}$ (right). The value of $n_{\sigma,\rm pred}$ is shown above each panel. Blue lines mark the reference filaments, while red lines indicate predicted filaments. Halos with $R_{200c} > 500$ ckpc are shown as magenta circles, and the $w_{\rm fil}$ gray-scale map is shown in the background.}
    \label{fig:res_mock}
\end{figure*}

\begin{table*}[htbp]
    \centering
    \caption{Optimal predicted thresholds $n_{\sigma,\rm pred,best}$ for $n_{\sigma,\rm ref}=2.5$ in each redshift bin, determined as the KDE peak of the $n_{\sigma,\rm pred,best}$ distribution.}
    {\renewcommand{\arraystretch}{1.3}
    \begin{tabular}{c|ccccc}
    \hline
        \multirow{2}{*}{$n_{\sigma,\rm ref}$} & \multicolumn{5}{c}{$n_{\sigma,\rm pred,best}$}\\
        \cline{2-6}
          & $0.50 < z < 0.82$ & $0.82 < z < 1.21$ & $1.21 < z < 1.91$ & $1.91 < z < 3.52$ & Entire Sample\\
        \hline
        2.5 & 1.53 & 1.61 & 1.56 & 1.86 & 1.74\\
        \hline
    \end{tabular}}
    \tablefoot{These values are adopted for extracting $\mathcal{F}_{\rm 2D,obs}$.}
    \label{tab:summary}
\end{table*}

\begin{figure*}[htbp]
\sidecaption
    \includegraphics[width=0.68\textwidth]{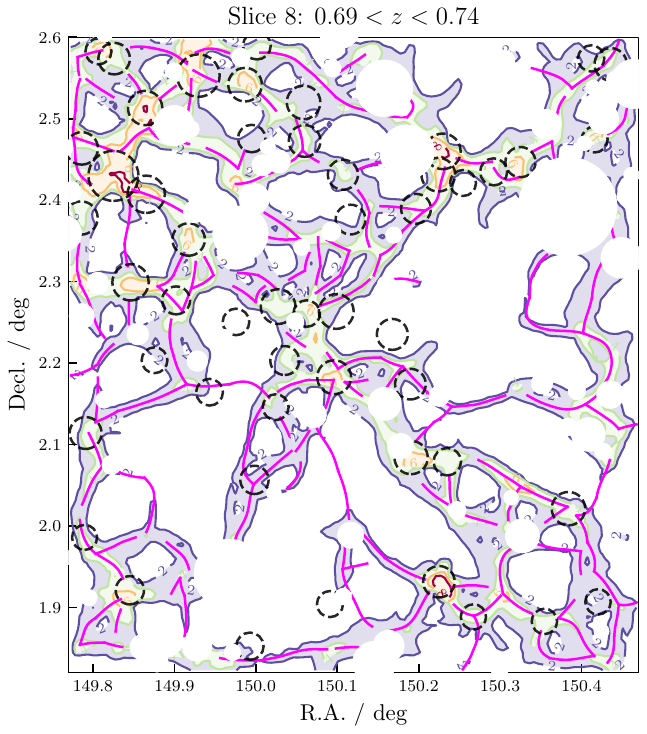}
    \caption{$w_{\rm fil}$ contour map of slice 8 ($0.69 < z < 0.74$) of the COSMOS-Web catalog, with $\ln(1+w_{\rm fil})$ contour levels from 2 to 8. Predicted filament segments extracted at the optimal $n_{\sigma,\rm pred}$ (magenta lines, Table~\ref{tab:summary}) are overlaid, together with galaxy groups (dashed black circles) from the overlapping slices 7, 8, and 9. Groups' $R_{200c}$ radii are shown to scale at the slice 8 mid-redshift ($z \approx 0.71$). HSC star masks are shown as opaque white filled regions.}
    \label{fig:res_obs}
\end{figure*}

Figure~\ref{fig:pred_opt_r50_nomw} shows the distributions of $n_{\sigma,\rm pred,best}$ and the corresponding $\Xi_{\rm best}$ values, obtained via kernel density estimation (KDE) for each redshift bin. The optimal $n_{\sigma,\rm pred}$ varies mildly with redshift (Table~\ref{tab:summary}), increasing from $n_{\sigma,\rm pred} \simeq 1.5$ in the lowest-redshift bin to $n_{\sigma,\rm pred} \simeq 1.9$ at $z > 1.9$. This trend reflects the decreasing galaxy number density and the correspondingly lower signal-to-noise ratio in the $w_{\rm fil}$ maps at higher redshifts, which requires a higher persistence threshold to suppress spurious structures. Over the full redshift range, the KDE peak of the $n_{\sigma,\rm pred,best}$ distribution is $n_{\sigma,\rm pred} \simeq 1.74$, with a corresponding $\Xi_{\rm best} \simeq 1.30$. The distance and angular separation statistics at the optimal thresholds (Figures~\ref{fig:dist_dis} and \ref{fig:ang_dis}) confirm that $\mathcal{F}_{\rm 2D,mock}$ faithfully reproduces $\mathcal{F}_{\rm 3D,ref}$. These metrics quantify the positional and directional agreement between the two filament catalogs: the distance separation measures how far apart (in 3D comoving space) the nearest matching segments are, and the angular separation measures how well their local orientations align. We adopted the per-bin $n_{\sigma,\rm pred,best}$ values in Table~\ref{tab:summary} for extracting $\mathcal{F}_{\rm 2D,obs}$.

\section{Results}\label{sec:res}

\begin{figure*}[htbp]
    \includegraphics[width=0.97\textwidth]{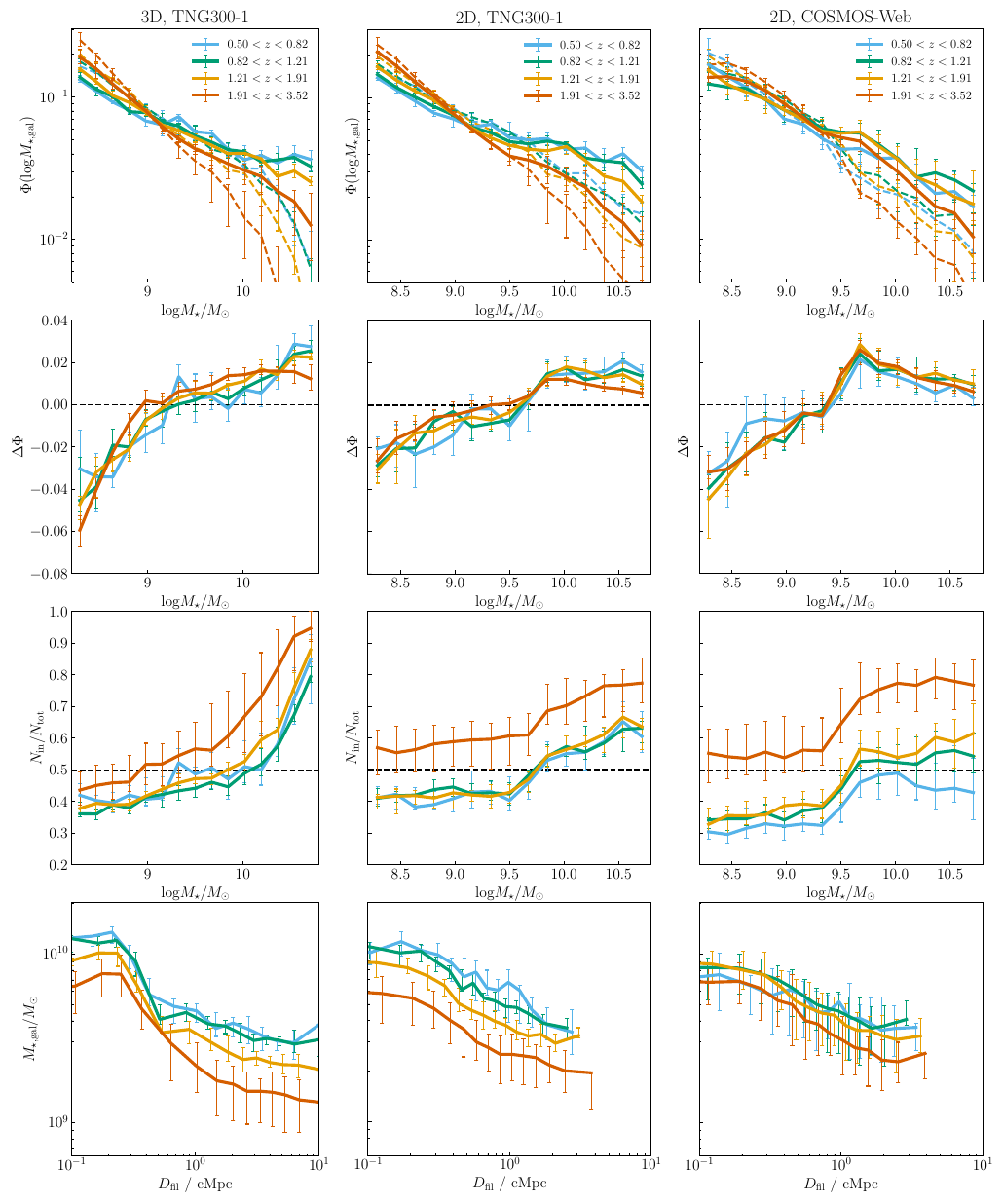}
    \caption{From top to bottom: normalized stellar mass functions (GSMF) for galaxies inside and outside filaments (solid and dashed lines, respectively); the difference $\Delta\Phi \equiv \Phi_{\rm in} - \Phi_{\rm out}$ between the inside- and outside-filament GSMFs; the number fraction of galaxies inside filaments $N_{\rm in}/N_{\rm tot}$ as a function of stellar mass; and the average galaxy stellar mass in bins of (projected) distance from the filament spine. Columns correspond to the 3D reference catalog ($\mathcal{F}_{\rm 3D,ref}$, left), the 2D mock catalog ($\mathcal{F}_{\rm 2D,mock}$, middle), and the 2D COSMOS-Web catalog ($\mathcal{F}_{\rm 2D,obs}$, right). Colored lines indicate the bin-averaged trends over all slices in each redshift bin. Error bars denote $1\sigma$ uncertainties.}
    \label{fig:astro_props}
\end{figure*}

\begin{figure*}
    \centering
    \includegraphics[width=\linewidth]{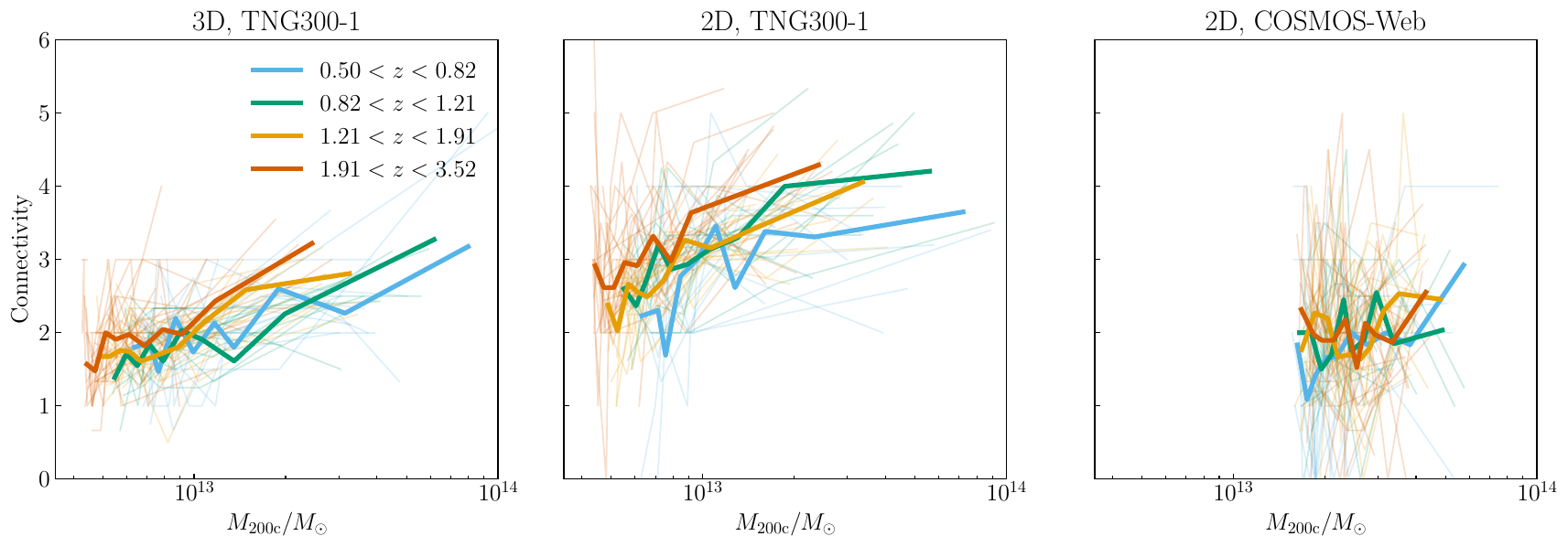}
    \caption{The average connectivity in bins of $M_{200c}$ for galaxy groups and clusters, shown for $\mathcal{F}_{\rm 3D,ref}$, $\mathcal{F}_{\rm 2D,mock}$, and $\mathcal{F}_{\rm 2D,obs}$. Thin lines show the results for each slice, while thick lines show the averaged connectivity.}
    \label{fig:conn}
\end{figure*}

We first present an overview of the filament catalog and $w_{\rm fil}$ maps (Section~\ref{sec:cat_overview}), then examine representative best- and worst-match extraction results in the mock catalog to assess whether the optimal $n_{\sigma,\rm pred}$ values yield coherent 3D structures (Section~\ref{sec:vis}). We then evaluate three astrophysical diagnostics---the galaxy stellar mass function inside and outside filaments, the stellar mass gradient toward the filament spine, and the connectivity--mass relation of galaxy groups and clusters---for $\mathcal{F}_{\rm 3D,ref}$, $\mathcal{F}_{\rm 2D,mock}$, and $\mathcal{F}_{\rm 2D,obs}$. This comparison tests how well the 2D reconstruction recovers the environmental trends encoded in the full 3D skeleton.

\subsection{Summary and visualization for extraction results}\label{sec:cat_overview}

Since each slice overlaps by half with the preceding one (except for the first slice), we report the extraction results for the even (slices 0, 2, 4, ..., 66) and odd (slices 1, 3, 5, ..., 65) samples separately. We excluded all filaments shorter than 1 cMpc, as these are likely to represent spurious detections. For the even sample, covering 0.50 < $z$ < 3.61, we identified 12,523 filaments, comprising a total of 136,210 segments. The per-segment geometry (endpoint coordinates, length, and robustness of all 136,210 segments) is released as a companion segment catalog alongside the filament catalog (Section~\ref{sec:conclusions}). Detailed information for each slice in the even sample is listed in Table~\ref{tab:info}. For brevity, the corresponding information for the odd sample is not shown here, but is available from the authors upon request.

For the astrophysical validation in Sections~\ref{sec:mass_gradient} and~\ref{sec:conn_mass}, we computed the projected distance from each galaxy (with $M_\star > 10^{8.2}\,M_\odot$) to its nearest filament. In all subsequent analyses, we included both the even and odd samples, as they are statistically indistinguishable and do not require separate analysis.

Figure~\ref{fig:res_mock} illustrates representative best- and worst-match extraction results for slice 8 ($0.69 < z < 0.74$) in $\mathcal{F}_{\rm 2D,mock}$, with $\mathcal{F}_{\rm 3D,ref}$ (blue and green lines for filaments below and above the 50\% robustness threshold, respectively) extracted at $n_{\sigma,\rm ref}=2.5$ and $\mathcal{F}_{\rm 2D,mock}$ (red lines) extracted at the $n_{\sigma,\rm pred}$ values that minimize and maximize $\Xi$ for this slice in each redshift bin. Galaxy groups with $R_{200c} > 500$ ckpc are shown as magenta circles.\label{sec:vis}

At the best-match $n_{\sigma,\rm pred}$, $\mathcal{F}_{\rm 2D,mock}$ traces both the main backbones and secondary branches of the projected $\mathcal{F}_{\rm 3D,ref}$ with acceptable fidelity, and most galaxy groups are connected to at least one filament in $\mathcal{F}_{\rm 2D,mock}$. At the worst-match $n_{\sigma,\rm pred}$ (typically the extreme values of the scan range), $\mathcal{F}_{\rm 2D,mock}$ either fragments into isolated, noise-dominated segments (at low $n_{\sigma,\rm pred}$) or is over-pruned and fails to capture the reference structure (at high $n_{\sigma,\rm pred}$). This confirms that the $\Xi$ minimization procedure in Section~\ref{sec:optimize} selects $n_{\sigma,\rm pred}$ values that correspond to visually coherent filament maps.

For $\mathcal{F}_{\rm 2D,obs}$, Figure~\ref{fig:res_obs} shows the filament probability ($w_{\rm fil}$) contour map of slice 8 ($0.69 < z < 0.74$), with $\ln(1+w_{\rm fil})$ levels from 2 to 8. Filaments in $\mathcal{F}_{\rm 2D,obs}$ extracted at the optimal $n_{\sigma,\rm pred}$ (blue lines, Table~\ref{tab:summary}) trace the ridges of the probability map. Galaxy groups from \cite{Toni2025} in slices 7, 8, and 9 are overlaid as dashed black circles, with $R_{200c}$ radii drawn to scale at the slice 8 mid-redshift ($z \approx 0.71$). HSC star masks are shown as opaque white filled regions. The $w_{\rm fil}$ map reveals a prominent filamentary concentration in the upper-left corner of the field, where massive galaxies and large groups are clustered, and two under-dense void regions adjacent to a filamentary ridge extending toward the lower-right corner. All galaxy groups are connected to at least one predicted filament and lie near regions of elevated $w_{\rm fil}$, confirming that the extraction method produces filament networks that are both statistically robust and qualitatively consistent with the observed group distribution.

\subsection{Validation for extraction results}

In the following, we used astrophysical diagnostics, specifically galaxy properties within the cosmic web and the connectivity--mass relation, to test how well $\mathcal{F}_{\rm 2D,mock}$ and $\mathcal{F}_{\rm 2D,obs}$ reproduce the environmental trends seen in $\mathcal{F}_{\rm 3D,ref}$ and in observations. Throughout this section, the stellar-mass threshold for galaxies is set to $10^{8.2}\,M_\odot$ for both $\mathcal{F}_{\rm 2D,mock}$ and $\mathcal{F}_{\rm 2D,obs}$, corresponding to the mass-complete sample of COSMOS-Web DR1 up to redshift $z \sim 3.5$. Note that the stellar particle mass resolution in TNG300-1 is $\sim 1.1 \times 10^7\,M_\odot$, implying that our sample includes galaxies resolved with as few as $\sim 20$ stellar particles. Such marginally resolved systems may introduce systematic bias at the low-mass end ($10^{8.2}\,M_\odot < M_\star < 10^9\,M_\odot$); nevertheless, we expect this effect to be negligible for the primary validation results presented below.

\subsubsection{Galaxy stellar mass gradient toward the filament spine}\label{sec:mass_gradient}

Figure~\ref{fig:astro_props} presents, from top to bottom, the galaxy stellar mass function (GSMF) inside and outside filaments, the difference $\Delta\Phi \equiv \Phi_{\rm in} - \Phi_{\rm out}$ between the inside- and outside-filament GSMFs, the fraction of galaxies in filaments ($N_{\rm in}/N_{\rm tot}$), and the mean stellar mass as a function of distance to the filament spine, shown for four redshift bins and for the $\mathcal{F}_{\rm 3D,ref}$, $\mathcal{F}_{\rm 2D,mock}$, and $\mathcal{F}_{\rm 2D,obs}$ catalogs. We adopted the redshift-dependent filament radius $R_{\rm fil}$ from Figure~3 of \cite{Wang2024} and fit a fourth-order polynomial to their measurements to determine the radius in each slice. Table~\ref{tab:rfil} lists the range of $R_{\rm fil}$ values spanned within each redshift bin, computed from the polynomial fit at the bin edges. For the 2D catalogs ($\mathcal{F}_{\rm 2D,mock}$ and $\mathcal{F}_{\rm 2D,obs}$), we applied a scaling factor of 0.6 to $R_{\rm fil}$ to account for the fact that projected distances systematically underestimate the true 3D galaxy--filament separation; this correction is calibrated to bring the 2D mock $N_{\rm in}/N_{\rm tot}$ into agreement with the 3D reference. A galaxy is classified as inside a filament if its perpendicular distance to the nearest filament segment is smaller than the filament radius of its slice.

\begin{table*}[htbp]
    \centering
    \caption{Adopted filament radius $R_{\rm fil}$ for the four redshift bins.}
    \label{tab:rfil}
    {\renewcommand{\arraystretch}{1.3}
    \begin{tabular}{lcccc}
    \hline
    Redshift bin & $0.50 < z < 0.82$ & $0.82 < z < 1.21$ & $1.21 < z < 1.91$ & $1.91 < z < 3.52$ \\
    \hline
    $z_{\rm mid}$ & 0.65 & 1.00 & 1.55 & 2.65 \\
    $R_{\rm fil}$ (3D) [cMpc] & 0.82--0.86 & 0.82--0.84 & 0.84--1.07 & 1.07--2.15 \\
    $R_{\rm fil}$ (2D) [cMpc] & 0.49--0.52 & 0.49--0.50 & 0.50--0.64 & 0.64--1.29 \\
    \hline
    \end{tabular}}
    \tablefoot{Taken from \cite{Wang2024} and interpolated via a fourth-order polynomial fit. Values are given as the range spanned within each bin, from the lower to upper redshift edge. The 2D catalogs use $0.6\,R_{\rm fil}$ to correct for line-of-sight projection.}
\end{table*}

Across all redshift bins, $\mathcal{F}_{\rm 3D,ref}$ exhibits an upturn in the normalized GSMF at the high-mass end ($\gtrsim 10^{10.0}\,M_\odot$) for galaxies inside filaments, and a deficit relative to galaxies outside filaments at the low-mass end ($\lesssim 10^{9.0}\,M_\odot$). This upturn grows more pronounced toward lower redshifts. $\mathcal{F}_{\rm 2D,mock}$ recovers both trends, indicating that projection from 3D to 2D preserves the preferential clustering of massive galaxies near filaments. $\mathcal{F}_{\rm 2D,obs}$ likewise shows an upturn at the high-mass end, although the redshift evolution seen in the mocks is not detected. In the COSMOS-Web data, the GSMFs inside and outside filaments diverge significantly below $M_\star \sim 10^{9.5}\,M_\odot$; the origin of this divergence remains unclear.

The second row of Figure~\ref{fig:astro_props} shows the difference $\Delta\Phi \equiv \Phi_{\rm in} - \Phi_{\rm out}$ between the inside- and outside-filament GSMFs. A positive $\Delta\Phi$ indicates an excess of galaxies inside filaments relative to outside at a given stellar mass. In all three catalogs, $\Delta\Phi$ crosses from negative to positive at $M_\star \sim 10^{9.5}\,M_\odot$, confirming that high-mass galaxies preferentially reside in filaments while low-mass galaxies are more abundant outside. For $\mathcal{F}_{\rm 3D,ref}$, the amplitude of $\Delta\Phi$ at the high-mass end grows toward lower redshift, reflecting the build-up of massive galaxies in the filamentary environment over cosmic time. This trend is broadly recovered by $\mathcal{F}_{\rm 2D,mock}$, and is marginally detected in the COSMOS-Web data, where the large Poisson uncertainties at the high-mass end limit the significance.

The third row of Figure~\ref{fig:astro_props} shows $N_{\rm in}/N_{\rm tot}$ as a function of stellar mass. In all three catalogs, the fraction of galaxies inside filaments rises with stellar mass, confirming that the GSMF upturn at the high-mass end is driven by the greater abundance of massive galaxies in filaments. Following \cite{Wang2024}, the characteristic filament radius increases with redshift at $z \gtrsim 1.0$, which explains the systematically higher $N_{\rm in}/N_{\rm tot}$ in the two earlier redshift bins ($1.21 < z < 1.91$). The upturn in $\mathcal{F}_{\rm 2D,mock}$ and $\mathcal{F}_{\rm 2D,obs}$ is weaker than in $\mathcal{F}_{\rm 3D,ref}$, because projected distances systematically underestimate the true galaxy--filament separation, causing more low-mass galaxies to be misclassified as inside filaments. We note that the adopted $R_{\rm fil}$ values represent the redshift-dependent mean filament radius; in reality, filament radius correlates with filament length \citep{Wang2024}, which our single-radius-per-slice approach averages over. As a robustness check, we have verified that using a fixed filament radius ($R_{\rm fil} = 0.5$ cMpc for all slices) yields qualitatively consistent results for all three catalogs, indicating that the conclusions are not sensitive to the precise choice of $R_{\rm fil}$.

We further tested the reliability of the reconstructed filaments by examining the stellar-mass gradient toward the filament spine. The bottom row of Figure~\ref{fig:astro_props} shows the mass-distance relation in four redshift bins. In $\mathcal{F}_{\rm 3D,ref}$, massive galaxies are concentrated near the filament spine: the mean stellar mass rises steeply within $D_{\rm fil} \lesssim 0.4$ cMpc and flattens beyond $D_{\rm fil} \gtrsim 1.0$ cMpc. $\mathcal{F}_{\rm 2D,mock}$ and $\mathcal{F}_{\rm 2D,obs}$ recover a weaker but detectable version of this trend, indicating that line-of-sight projection does not erase the environmental signal. However, the sharp rise seen at $D_{\rm fil} \sim 0.4$ cMpc in 3D is not recovered in 2D. In both $\mathcal{F}_{\rm 3D,ref}$ and $\mathcal{F}_{\rm 2D,mock}$, the average stellar mass increases toward lower redshifts across the full range of $D_{\rm fil}$, reflecting galaxy mass growth over cosmic time. Although this redshift trend is less pronounced in $\mathcal{F}_{\rm 2D,obs}$, the close agreement between $\mathcal{F}_{\rm 2D,obs}$ and $\mathcal{F}_{\rm 2D,mock}$ shows that the optimized thresholds yield a 2D skeleton that reliably traces the underlying 3D galaxy--filament connection, so that environmental trends measured from COSMOS-Web can be physically interpreted with the TNG300-1 simulation.

\subsubsection{Connectivity-mass relation of galaxy groups and clusters}\label{sec:conn_mass}

Connectivity \citep{Codis2018} is defined as the number of filaments connected to a given halo; more massive halos are typically connected to more filaments. For $\mathcal{F}_{\rm 3D,ref}$, we measured connectivity as the number of filament segments that intersect a sphere of radius $R=1.5\,R_{200c}$ centered on the halo, following \cite{Darragh2019}. The COSMOS-Web group sample is from \cite{Toni2025} (Section~\ref{sec:obs_data}).

Figure~\ref{fig:conn} presents the connectivity--mass relation for $\mathcal{F}_{\rm 3D,ref}$, $\mathcal{F}_{\rm 2D,mock}$, and $\mathcal{F}_{\rm 2D,obs}$. We binned systems by equal numbers per bin. The COSMOS-Web groups from \cite{Toni2025} span $M_{200c} \sim (5\times10^{12}, \text{a few}\times10^{13})\,M_\odot$, while the TNG300-1 mock clusters extend to higher masses ($\lesssim 10^{14}\,M_\odot$). In the $M_{200c}$ range where the two datasets overlap, the connectivity--mass trends are consistent between the COSMOS-Web and TNG mock samples. At higher masses probed only by the mocks, connectivity continues to rise with $M_{200c}$, producing the overall positive correlation. At low redshift ($0.5 < z < 0.82$, blue lines), the mean connectivity is 2--4, consistent with previous studies \citep{Darragh2019,Yu2025}. The $\mathcal{F}_{\rm 2D,obs}$ recovers this positive trend at the group-mass scale, with typical low redshift connectivity of 1--3. The $\mathcal{F}_{\rm 2D,mock}$ connectivity is systematically higher than in $\mathcal{F}_{\rm 3D,ref}$, because projection raises the filament density on slices while the halo density remains fixed. Taken together, these results show that $\mathcal{F}_{\rm 2D,mock}$ and $\mathcal{F}_{\rm 2D,obs}$ recover the qualitative connectivity--mass relation seen in $\mathcal{F}_{\rm 3D,ref}$, supporting the reliability of the extraction method.

\subsection{Probability calibration of the \texorpdfstring{$w_{\rm fil}$}{w\_fil} map}\label{sec:prob_calib}

The $w_{\rm fil}$ map records, at each pixel, the total comoving length of filament segments that intersect that pixel across all 500 redshift realizations. Although $w_{\rm fil}$ is not a formal probability, its magnitude should correlate with the likelihood that a true 3D filament passes through that location. To verify this, we calibrated the $w_{\rm fil}$ values against $\mathcal{F}_{\rm 3D,ref}$ in the mock data.

For each mock slice, we rasterized $\mathcal{F}_{\rm 3D,ref}$ (extracted at $n_{\sigma,\rm ref}=2.5$) onto the same regular grid as the $w_{\rm fil}$ map, producing a binary ``hit'' mask. We then binned all non-zero pixels by their $w_{\rm fil}$-map value $x \equiv \ln(1+w_{\rm fil})$ in steps of $0.5$ and computed the fraction of pixels in each bin that are intersected by a reference filament---the reference hit-rate. This is performed separately for the four redshift bins defined in Section~\ref{sec:tomo_slice}.

\begin{figure}[htbp]
    \centering
    \includegraphics[width=0.48\textwidth]{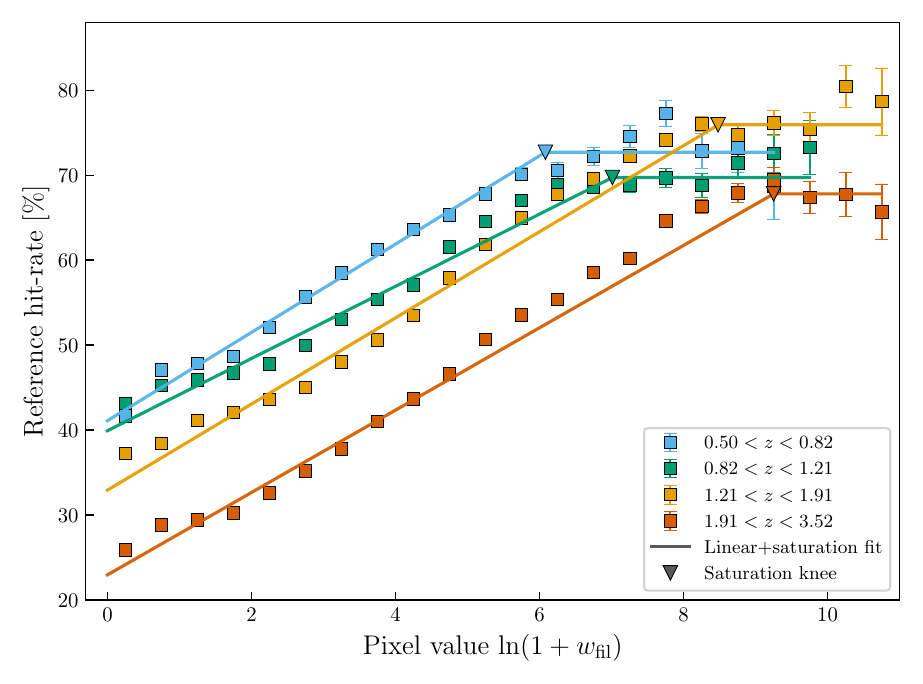}
    \caption{Calibration of $w_{\rm fil}$ map pixel values against the reference filament hit-rate for the four redshift bins. The 3D reference filaments ($\mathcal{F}_{\rm 3D,ref}$) are rasterized onto the $w_{\rm fil}$ grid, and the fraction of pixels intersected by a reference filament (hit-rate) is plotted as a function of the stored pixel value $\ln(1+w_{\rm fil})$. {Solid lines show the per-bin linear$+$saturation fit (Appendix~\ref{app:wfil_atlas}, Table~\ref{tab:probfit}), and downward triangles mark the saturation knee $x_{\rm sat}$.} Error bars show the binomial uncertainty in each bin. Only bins with at least 100 pixels are shown.}
    \label{fig:prob_calib}
\end{figure}

Figure~\ref{fig:prob_calib} shows the resulting calibration curves. Across all four redshift bins, the reference hit-rate rises monotonically with the pixel value $\ln(1+w_{\rm fil})$, from $\sim$25--40\% near zero to $\sim$70--80\% at $\ln(1+w_{\rm fil}) \gtrsim 8$, before flattening to a saturation plateau. The Spearman rank correlation is highly significant ($\rho \approx 0.94$--$0.97$, $p \ll 0.001$) in all bins, confirming a strong monotonic relationship. We modeled each bin's calibration as a linear rise followed by a saturation plateau, and the fitted curves (solid lines in Figure~\ref{fig:prob_calib}) reproduce the measured hit-rates well; the mapping and its per-bin parameters are given in Appendix~\ref{app:wfil_atlas}. The calibration curves are broadly similar across redshift, with a mild trend towards lower intercepts and later saturation at higher $z$, so that the mapping between $w_{\rm fil}$ and the likelihood of tracing a true 3D filament evolves only weakly. This result validates the quantitative interpretation of the $w_{\rm fil}$ map: higher pixel values correspond to a higher probability of genuine filamentary structure, and the contour levels shown in Figure~\ref{fig:res_obs} can be interpreted as graduated confidence intervals for the presence of a filament. As a companion data product to the filament catalog, we released the calibrated filament-probability maps for all 67 slices (Appendix~\ref{app:wfil_atlas}), which give the per-pixel probability of tracing a true 3D filament and can be used to evaluate the likelihood of filamentary structure at any sky position within a given redshift interval.

\section{Conclusions and discussions} \label{sec:conclusions}

\begin{figure*}
    \centering
    \includegraphics[width=1.0\textwidth]{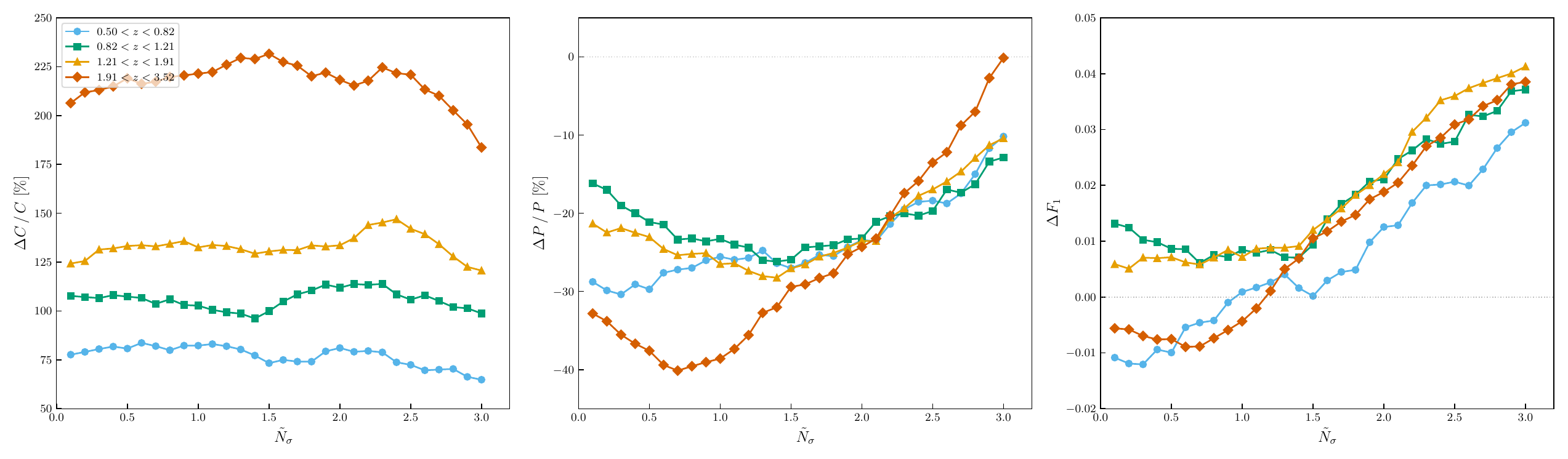}
    \caption{Comparison of the stacked PDF($z$) method ($\mathcal{F}_{\rm 2D,mock}$) and the conventional single-$z$ method ($\mathcal{F}_{\rm single}$) using completeness ($C$), purity ($P$), and $F_1$ score, evaluated against the 3D reference skeleton ($\mathcal{F}_{\rm 3D,ref}$) at $n_{\sigma,\rm ref}=2.5$. Left: Relative completeness difference $\Delta C/C$ as a function of the predicted extraction threshold $n_{\sigma,\rm pred}$. Middle: Relative purity difference $\Delta P/P$. Right: Absolute $F_1$-score difference $\Delta F_1 = F_{1,\rm stack} - F_{1,\rm single}$. Colored curves correspond to the four redshift bins indicated in the legend. The stacked method achieves 50--250\% higher completeness at the cost of a 10--40\% lower purity, yielding a net positive $\Delta F_1 \approx 0.01$--$0.03$ across all bins.}
    \label{fig:comp_q50}
\end{figure*}

This work presents a method to extract projected 2D filaments from tomographic slices in the presence of photometric redshift uncertainties, extending the approach of \cite{EuclidCollab2025} by exploiting galaxy PDF(z) information. The optimal extraction parameters were calibrated through mock experiments: we fixed $\mathcal{F}_{\rm 3D,ref}$ at $n_{\sigma,\rm ref}=2.5$ and scanned $n_{\sigma,\rm pred}$ to find the values that minimize the filament-similarity metric $\Xi$ between $\mathcal{F}_{\rm 3D,ref}$ and $\mathcal{F}_{\rm 2D,mock}$ extracted from $w_{\rm fil}$ maps. We validated the resulting catalog and $w_{\rm fil}$ maps by comparing spatial and angular separations between $\mathcal{F}_{\rm 2D,mock}$ and $\mathcal{F}_{\rm 3D,ref}$, and by testing whether $\mathcal{F}_{\rm 2D,mock}$ and $\mathcal{F}_{\rm 2D,obs}$ recover key astrophysical trends, including galaxy mass gradients and the connectivity--mass relation, in both mock and observational data.

\subsection{Limitations and potential caveats}

First, the calibration of DisPerSE persistence thresholds is simulation-driven. We optimized $n_{\sigma,\rm pred}$ by maximizing 3D--2D agreement in a TNG300-1 light cone, using the grid-based $w_{\rm fil}$ maps and the \texttt{-cut} option with an ECDF-based $n_\sigma{\rightarrow}\texttt{cut}$ mapping. The statistical properties of these fields differ from the Poisson point processes for which DisPerSE's significance levels were originally defined, and we do not attempt a full noise or false-discovery analysis for the grid representation. Our thresholds should therefore be regarded as empirically optimal for TNG300-1, rather than as universal significance cuts. If the real Universe differs from TNG300-1 in its web topology, galaxy bias, or baryonic effects, our catalog may be somewhat over- or under-pruned.

Second, our forward modeling of photometric redshift PDFs in the mock light cone is approximate. We reproduce the observed $\Delta z(z, m_{\rm Ks})$ distribution (Section~\ref{sec:optimize}), but we model individual PDF($z$) as simple Gaussians and do not capture strongly non-Gaussian or multi-modal PDFs. Catastrophic outliers and asymmetric tails could alter how galaxies scatter between slices and, in turn, affect filament completeness and purity. Nevertheless, the mock catalog reproduces the observed $\Delta z$--$z$ and $\Delta z$--$m_{\rm Ks}$ relations to within $1\sigma$ (Figures~\ref{fig:zerr_obs} and~\ref{fig:zerr_check}), indicating that the Gaussian approximation is sufficient at the statistical level for the filament extraction and validation performed in this work. A more faithful treatment using realistic PDF($z$) templates matched to individual galaxies is deferred to future work.

Finally, our validation focuses on a limited set of diagnostics: stellar mass functions in and out of filaments, mass gradients toward filaments, and the connectivity--mass relation. These tests demonstrate that our 2D catalog and $w_{\rm fil}$ maps encode meaningful LSS information, but they do not exhaust all possible uses of the catalog. We have not yet examined, for example, star-formation, color, morphology, or AGN trends with filament environment, nor have we systematically analyzed complementary web components such as sheets and voids. Users of the catalog should therefore regard our current results as a proof of concept, and we encourage future work to stress-test the catalog under alternative simulations, parameter choices, and a broader set of observables.

\subsection{Comparison with conventional extraction methods}\label{sec:comp_single}

Filament extraction is method-dependent \citep{Libeskind2018}, and different techniques can yield different skeletons. We therefore compared our stacked PDF(z) approach ($\mathcal{F}_{\rm 2D,mock}$ and $\mathcal{F}_{\rm 2D,obs}$) with the conventional single-$z$ method ($\mathcal{F}_{\rm single}$), in which DisPerSE is applied directly to one redshift realization per slice.

Figure~\ref{fig:comp_q50} quantifies this comparison using completeness ($C$), purity ($P$), and $F_1$ score, evaluated against $\mathcal{F}_{\rm 3D,ref}$ at $n_{\sigma,\rm ref}=2.5$. The three panels show, respectively, the relative completeness difference $\Delta C/C$, the relative purity difference $\Delta P/P$, and the absolute $F_1$-score difference $\Delta F_1 = F_{1,\rm stack} - F_{1,\rm single}$, all as functions of $n_{\sigma,\rm pred}$. Across all four redshift bins, $\mathcal{F}_{\rm 2D,mock}$ achieves 50--250\% higher completeness than $\mathcal{F}_{\rm single}$, at the cost of a 10--40\% lower purity. The net $F_1$ score is nevertheless positive ($\Delta F_1 \approx 0.01$--$0.03$), confirming that the stacked method yields an overall improvement in filament recovery. The advantage is most pronounced at intermediate thresholds ($n_{\sigma,\rm pred} \approx 1.0$--$2.0$) and in the lower-redshift bins, where the photometric redshift uncertainties are smaller and the PDF($z$) stacking more effectively suppresses noise.

The data products accompanying this paper consist of three components. The first is the COSMOS-Web filament catalog ($\mathcal{F}_{\rm 2D,obs}$), which contains 12,523 filaments across 67 tomographic slices covering $0.5 < z < 3.5$. For each filament, the catalog provides its sky coordinates, total comoving length, number of constituent segments, mean DisPerSE robustness, and redshift interval, enabling direct application to studies of galaxy--filament interactions, environmental quenching, and LSS topology. The second is the accompanying segment catalog, which resolves these filaments into their 136,210 constituent segments. The third is the set of 67 calibrated filament-probability maps, one per tomographic slice, which give the per-pixel probability ($P \in [0,1]$) that a location traces a true 3D filament, obtained by applying our probability calibration (Appendix~\ref{app:wfil_atlas}) to the stacked PDF($z$) realizations. These maps allow users to evaluate the filament probability at any sky position within a given redshift slice. Together, the filament and segment catalogs and the probability maps constitute a versatile resource for investigating the cosmic web and its role in galaxy evolution across cosmic time.

Despite limitations arising from $z_{\rm phot}$ uncertainties, projection effects, and the method dependence of filament extraction, the data products presented in this work provide a foundation for broader studies of the cosmic web across cosmic time. Future JWST multi-band programs, including COSMOS-3D, are expected to improve the accuracy of high-redshift filament detection, enabling a more reliable 3D reconstruction of the cosmic web before and around cosmic noon. We are currently investigating additional galaxy astrophysical properties in relation to filaments, as well as the redshift evolution of filamentary structures themselves. These studies will advance our understanding of galaxy-environment coevolution, particularly at $z \gtrsim 2$.

\section*{Data availability}                      

The filament catalog and the segment catalog are only available in electronic form at the CDS via anonymous ftp to cdsarc.u-strasbg.fr (130.79.128.5) or via http://cdsweb.u-strasbg.fr/cgi-bin/qcat?J/A+A/. The 67 filament-probability maps are likewise distributed in electronic form at the CDS.

\begin{acknowledgements}

This work was supported by National Natural Science Foundation of China (Grant No.12525302 and 12141301), Natural Science Foundation of Jiangsu Higher Education Institutions of China(Grant No. BK20250001), National Key R\&D Program of China (Grant no. 2023YFA1605600), Scientific Research Innovation Capability Support Project for Young Faculty (Project No. ZYGXQNJSKYCXNLZCXM-P3), the Fundamental Research Funds for the Central Universities with Grant no.KG202502, and the China Manned Space Program with grant no. CMS-CSST-2025-A04. We are grateful to Clotilde Laigle for contributions to this paper and for helpful suggestions and references regarding methods for incorporating redshift uncertainties. P.W. acknowledges the financial support from the NSFC (No.12595312, No.12473009), and is also sponsored by Shanghai Rising-Star Program (No. 24QA2711100). P.W. acknowledges supports from the China Manned Space Program with grant No.CMS-CSST-2025-A03.

\end{acknowledgements}

\bibliography{reference}
\bibliographystyle{aa}

\clearpage
\begin{appendices}
\counterwithin{equation}{section}
\counterwithin{figure}{section}
\counterwithin{table}{section}

\section{Detailed information of tomographic slices}\label{app:detail}

In this section, we present a supplementary table (Table~\ref{tab:info}) that provides detailed information on the number of filaments and segments for each slice with an even index ($\rm ID_i$). We do not report results for slices with odd indices, as odd and even slices correspond to the same light cone, with the even slices forming a more complete light cone spanning the redshift range 0.5\textendash3.5. Results for the odd slices are available upon request from the authors.

Table~\ref{tab:info} lists the correspondence between the COSMOS-Web slice index $i$ ($\rm ID_i$) and the TNG300-1 snapshot index $j$ ($\rm ID_j$), the redshift interval of each slice ($z$), the sample size ($N_{\rm gal}$), the spectroscopic redshift completeness ($c_{\rm spec}$), the side length of the COSMOS-Web field of view ($L_{\rm fov}$), and the numbers of filaments and segments in $\mathcal{F}_{\rm 2D,obs}$ ($N_{\rm fil,obs}$ and $N_{\rm seg,obs}$), $\mathcal{F}_{\rm 3D,ref}$ ($N_{\rm fil,ref}$ and $N_{\rm seg,ref}$), and $\mathcal{F}_{\rm 2D,mock}$ ($N_{\rm fil,pred}$ and $N_{\rm seg,pred}$).

\begin{table*}[htbp]
    \centering
    \caption{Detailed information of the tomographic slices and number of filaments/segments in each slice.}
    {\renewcommand{\arraystretch}{1.3}
    \begin{tabular}{cccccccccc}
    \hline
       Redshift bin & $\rm ID_{i}$ & z & $N_{\rm gal}$ & $c_{\rm spec}$/\% & $L_{\rm fov}$/cMpc & $N_{\rm fil,obs}$/$N_{\rm seg,obs}$ & $\rm ID_{j}$ & $N_{\rm fil,ref}$/$N_{\rm seg,ref}$ & $N_{\rm fil,pred}$/$N_{\rm seg,pred}$ \\
       \hline
        \multirow{7}{*}{(0.50, 0.82)} & 0 & (0.500, 0.545) & 490 & 56.1 & 23.7 & 111/880 & 68 & 138/3136 & 114/830 \\
        & 2 & (0.545, 0.592) & 160 & 60.6 & 25.5 & 110/898 & 66 & 163/4294 & 156/1114 \\
        & 4 & (0.592, 0.640) & 622 & 52.7 & 27.4 & 149/1235 & 64 & 198/4878 & 186/1404 \\
        & 6 & (0.640, 0.689) & 887 & 55.0 & 29.2 & 177/1382 & 62 & 184/5359 & 184/1415 \\
        & 8 & (0.689, 0.740) & 1248 & 56.2 & 31.0 & 168/1187 & 60 & 210/5443 & 218/1582 \\
        & 10 & (0.740, 0.792) & 418 & 56.9 & 32.9 & 162/1518 & 59 & 228/5779 & 207/1587 \\
        & 12 & (0.792, 0.846) & 761 & 42.0 & 34.7 & 178/1661 & 57 & 254/6872 & 265/2051 \\
        \hline
        \multirow{6}{*}{(0.82, 1.21)} & 14 & (0.846, 0.902) & 1683 & 36.4 & 36.5 & 274/1924 & 55 & 299/7584 & 343/2258 \\
        & 16 & (0.902, 0.960) & 1920 & 35.5 & 38.4 & 306/2401 & 53 & 310/7944 & 370/2805 \\
        & 18 & (0.960, 1.020) & 1797 & 22.0 & 40.2 & 367/2917 & 52 & 324/8942 & 391/3004 \\
        & 20 & (1.020, 1.082) & 889 & 19.3 & 42.0 & 311/2825 & 50 & 388/10415 & 447/3321 \\
        & 22 & (1.082, 1.146) & 670 & 14.3 & 43.8 & 255/2719 & 49 & 381/10307 & 491/3680 \\
        & 24 & (1.146, 1.213) & 1688 & 19.2 & 45.7 & 416/3354 & 47 & 438/11738 & 524/3823 \\
        \hline
        \multirow{8}{*}{(1.21, 1.91)} & 26 & (1.213, 1.283) & 1111 & 16.1 & 47.5 & 337/3239 & 46 & 457/13124 & 479/3695 \\
        & 28 & (1.283, 1.355) & 1055 & 15.5 & 49.3 & 418/3919 & 44 & 472/12582 & 505/3834 \\
        & 30 & (1.355, 1.430) & 1166 & 13.9 & 51.2 & 418/4053 & 43 & 479/11318 & 549/4224 \\
        & 32 & (1.430, 1.509) & 1067 & 19.9 & 53.0 & 326/3546 & 41 & 472/12785 & 561/4584 \\
        & 34 & (1.509, 1.590) & 755 & 12.6 & 54.8 & 352/4043 & 40 & 539/15915 & 610/4752 \\
        & 36 & (1.590, 1.675) & 2847 & 7.8 & 56.6 & 661/4937 & 39 & 559/14496 & 686/4882 \\
        & 38 & (1.675, 1.765) & 1499 & 7.6 & 58.5 & 543/4838 & 38 & 600/16046 & 711/5098 \\
        & 40 & (1.765, 1.858) & 1074 & 3.8 & 60.3 & 514/5502 & 36 & 645/19400 & 761/6201 \\
        \hline
        \multirow{12}{*}{(1.91, 3.52)} & 42 & (1.858, 1.955) & 875 & 4.6 & 62.1 & 506/5025 & 35 & 638/17121 & 819/5728 \\
        & 44 & (1.955, 2.057) & 1099 & 5.8 & 64.0 & 326/4201 & 34 & 662/19991 & 900/6413 \\
        & 46 & (2.057, 2.164) & 1326 & 6.6 & 65.8 & 381/4989 & 33 & 694/20402 & 889/7017 \\
        & 48 & (2.164, 2.277) & 883 & 8.6 & 67.6 & 359/4549 & 32 & 735/21332 & 989/7007 \\
        & 50 & (2.277, 2.395) & 1015 & 7.4 & 69.5 & 292/4426 & 31 & 759/23594 & 967/7783 \\
        & 52 & (2.395, 2.519) & 1950 & 6.4 & 71.3 & 580/6726 & 30 & 795/24375 & 1004/8145 \\
        & 54 & (2.519, 2.650) & 886 & 6.2 & 73.1 & 519/6453 & 29 & 764/25297 & 1145/9251 \\
        & 56 & (2.650, 2.788) & 792 & 6.3 & 74.9 & 298/4989 & 28 & 911/29448 & 1096/9041 \\
        & 58 & (2.788, 2.934) & 1019 & 5.7 & 76.8 & 462/6350 & 27 & 913/28442 & 1152/9067 \\
        & 60 & (2.934, 3.088) & 1223 & 4.7 & 78.6 & 521/6925 & 26 & 878/27672 & 1218/10085 \\
        & 62 & (3.088, 3.251) & 915 & 4.4 & 80.4 & 582/7535 & 25 & 939/29919 & 1150/9878 \\
        & 64 & (3.251, 3.423) & 769 & 3.9 & 82.3 & 421/6208 & 24 & 1050/34308 & 1255/10646 \\
        & 66 & (3.423, 3.606) & 739 & 2.4 & 84.1 & 723/8856 & 24 & 937/29950 & 1240/10602 \\
        \hline
    \end{tabular}}
    \label{tab:info}
\end{table*}

\section{Details of the threshold scan and similarity metric}\label{app:thresh_detail}

\subsection{Threshold scan setup and the \texorpdfstring{$n_{\sigma}{\rightarrow}\texttt{cut}$}{n\_sigma->cut} mapping}

For each slice in the mock catalog, $\mathcal{F}_{\rm 3D,ref}$ is extracted by running DisPerSE on the DTFE-computed density field of galaxy positions (see Section~\ref{sec:optimize}), using the \texttt{-nsig} parameter with $n_{\sigma,\rm ref} = 2.5$. This corresponds to a noise probability $p = \frac{1}{2}\,\mathrm{erfc}(2.5/\sqrt{2}) \approx 0.6\%$: only structures whose persistence exceeds the 99.4th percentile of the Poisson noise expectation are retained in $\mathcal{F}_{\rm 3D,ref}$.

For $\mathcal{F}_{\rm 2D,mock}$, the $w_{\rm fil}$ maps are defined on regular grids, and DisPerSE requires the \texttt{-cut} parameter (an absolute persistence threshold) rather than \texttt{-nsig}. To link the two, we constructed an $n_{\sigma}{\rightarrow}\texttt{cut}$ mapping for each slice:

\begin{enumerate}
    \item Generate 500 random-background realizations of the $w_{\rm fil}$ map by randomizing the sky positions of galaxies before stacking the DisPerSE skeletons; this preserves the number density and mask geometry while destroying any intrinsic LSS.
    \item Extract all persistence pairs (maxima--saddle point) from the random-background $w_{\rm fil}$ maps without applying any threshold, and pool the persistence values to build the empirical cumulative distribution function $F_{\rm rand}(r) = P(\mathrm{persistence} \leq r)$ for that slice.
    \item For a target $n_{\sigma}$, compute the noise probability $p = \frac{1}{2}\,\mathrm{erfc}(n_{\sigma}/\sqrt{2})$, and set the \texttt{-cut} value to $F_{\rm rand}^{-1}(1-p)$---the persistence threshold above which only a fraction $p$ of random-background pairs would survive.
\end{enumerate}

This procedure ensures that $n_{\sigma,\rm pred}$ has the same statistical interpretation (fraction of Poisson-noise fluctuations) when applied to the $w_{\rm fil}$ maps via \texttt{-cut}, as $n_{\sigma,\rm ref}$ has when applied to the galaxy distribution via \texttt{-nsig}.

We scanned $n_{\sigma,\rm pred}$ from 0.1 to 3.0 in steps of 0.1, converted each value to the corresponding \texttt{-cut} via the above mapping, and extracted $\mathcal{F}_{\rm 2D,mock}$ from the $w_{\rm fil}$ map at that threshold. In the main analysis, we restricted $\mathcal{F}_{\rm 3D,ref}$ to the 50\% most robust filaments (ranked by the DisPerSE \texttt{-robustness} value), so that the optimization targets structures that are most reliably traceable in the observations. Results without this robustness cut are reported in Appendix~\ref{app:zero}.

\subsection{Similarity metric and implementation}

When quantifying similarity between $\mathcal{F}_{\rm 3D,ref}$ and $\mathcal{F}_{\rm 2D,mock}$, we adopted the algorithm of \cite{EuclidCollab2025}. For each filament segment in $\mathcal{F}_{\rm 2D,mock}$ ($\mathcal{F}_{\rm 3D,ref}$), the nearest segment in $\mathcal{F}_{\rm 3D,ref}$ ($\mathcal{F}_{\rm 2D,mock}$) is found to form a segment pair. The spatial separation between the midpoints of each pair, and the angular separation between the pair (restricted to $[0^\circ, 90^\circ]$), yield the distance and angular separation distributions between $\mathcal{F}_{\rm 3D,ref}$ and $\mathcal{F}_{\rm 2D,mock}$ in both directions (3D$\rightarrow$2D and 2D$\rightarrow$3D). These distributions are expressed as the fraction of segments within each distance or angle bin relative to the total number of segments in $\mathcal{F}_{\rm 2D,mock}$ ($\mathcal{F}_{\rm 3D,ref}$) for the 2D$\rightarrow$3D (3D$\rightarrow$2D) comparison. Based on these distributions, the metric $\Xi$ captures differences in the peak of the distance separation distribution, the dispersion of the angular distribution, and the shape variations in the residual distributions. A smaller $\Xi$ indicates higher similarity between the two filament sets.

\subsection{Expectation values and dispersions for distance/angle distributions}

In Figures~\ref{fig:dist_dis} and \ref{fig:ang_dis}, the expectation values and standard deviations of the distance and angular separation distributions are computed following \cite{EuclidCollab2025}:

\begin{equation}
    x_{\rm peak}=\Sigma x f(x),
\end{equation}

\begin{equation}
    x_{\rm dev}=\sqrt{\Sigma f(x)(x-x_{\rm peak})^2},
\end{equation}

\section{Results without robustness cuts}\label{app:zero}

\begin{figure*}
    \centering
    \includegraphics[width=1\linewidth]{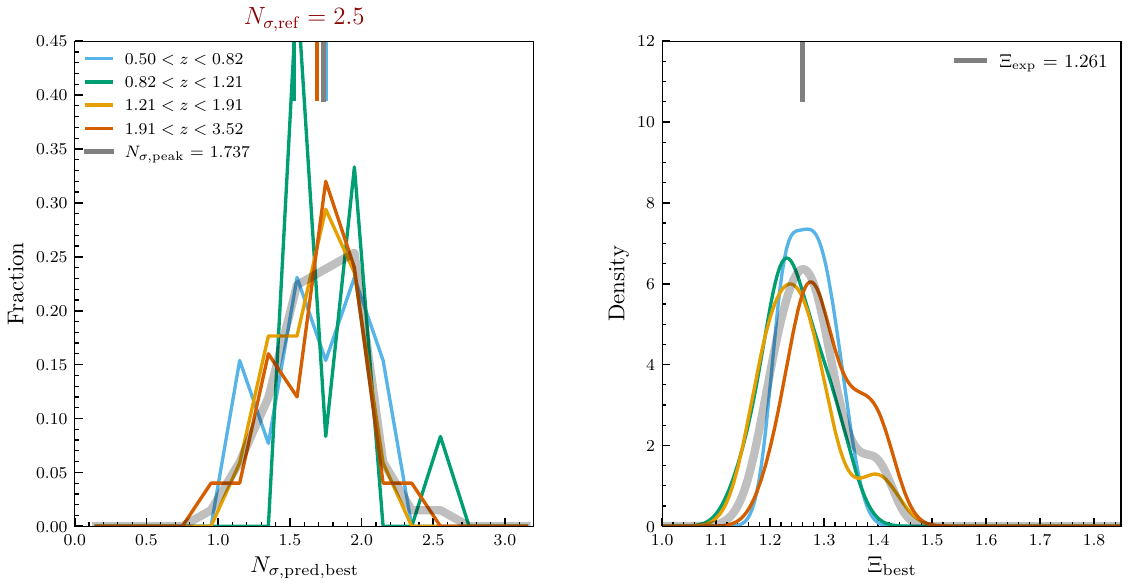}
    \caption{Similar to figure~\ref{fig:pred_opt_r50_nomw}, but without the 50\% robustness cut for $\mathcal{F}_{\rm 3D,ref}$.}
    \label{fig:pred_opt_r0_nomw}
\end{figure*}

Figure~\ref{fig:pred_opt_r0_nomw} presents the results of the threshold optimization before excluding $\mathcal{F}_{\rm 3D,ref}$ filaments with robustness values below the median.

\section{Probability mapping and calibrated data product}\label{app:wfil_atlas}

\begin{figure*}[htbp]
    \centering
    \includegraphics[width=0.8\textwidth]{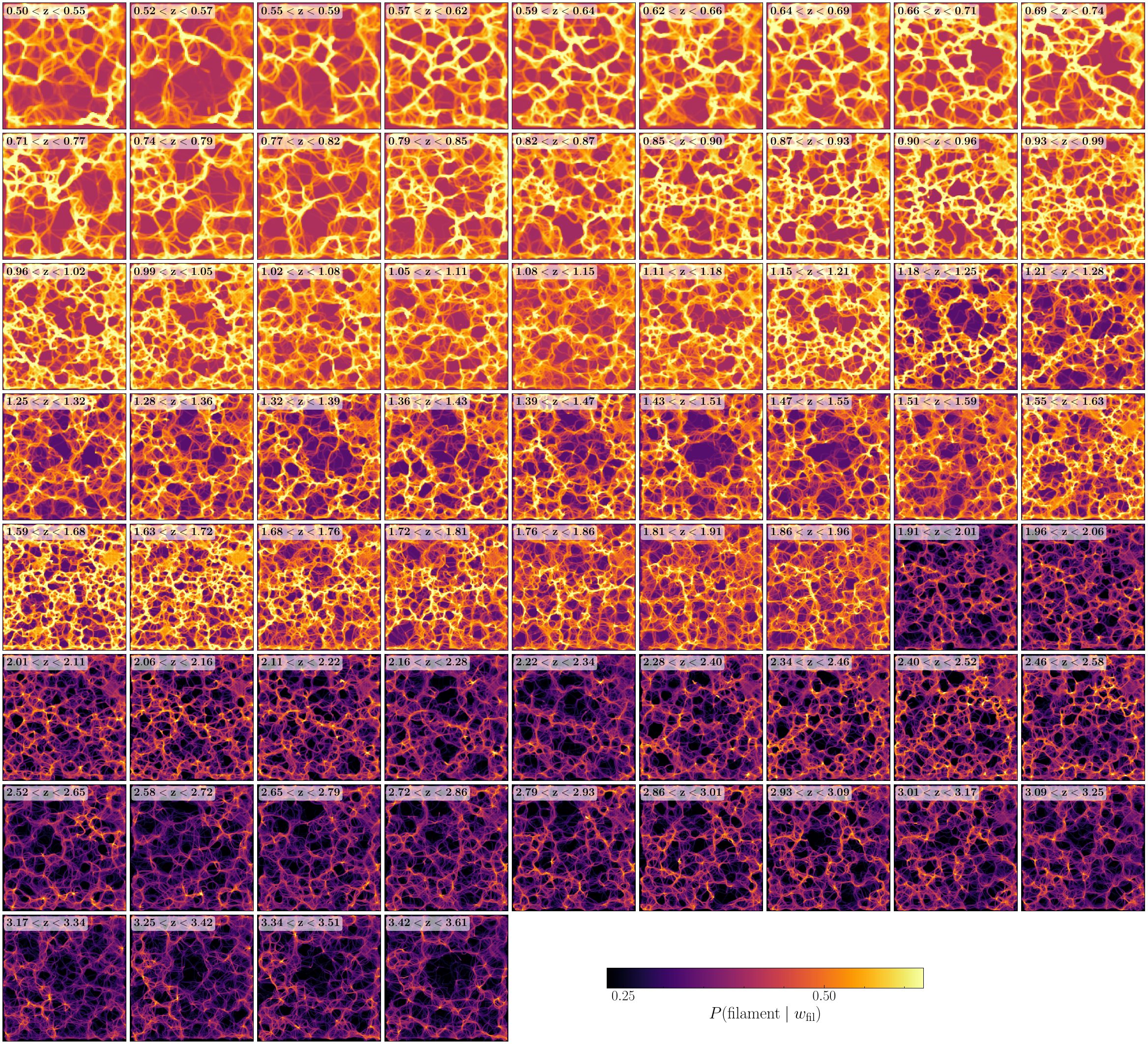}
    \caption{Calibrated filament-probability maps for all 67 tomographic slices. Each panel corresponds to one slice (slice~0 at top-left, slice~66 at bottom-right), showing the per-pixel filament probability $P \in [0,1]$ on the rotated COSMOS-Web footprint.}
    \label{fig:wfil_grid}
\end{figure*}

Figure~\ref{fig:wfil_grid} presents the released calibrated filament-probability maps for all 67 tomographic slices (slice~0 at top-left, slice~66 at bottom-right). Each panel shows the per-pixel filament probability $P \in [0,1]$ on the rotated COSMOS-Web footprint, obtained by applying the calibration described below to the $w_{\rm fil}$ map of each slice.

To turn the $w_{\rm fil}$ maps into a directly interpretable filament-probability product, we parametrized the reference hit-rate calibration of Section~\ref{sec:prob_calib} with an analytic mapping. In each redshift bin $b$, the hit-rate rises approximately linearly with the pixel value $x \equiv \ln(1+w_{\rm fil})$ and then flattens to a plateau. We therefore modeled the filament probability as a piecewise linear$+$saturation function,
\begin{equation}\label{eq:probcal}
    P_b(x) =
    \begin{cases}
        P_{0,b} + k_b\,x, & x < x_{{\rm sat},b},\\[4pt]
        P_{{\rm sat},b}, & x \ge x_{{\rm sat},b},
    \end{cases}
\end{equation}
where $P_{{\rm sat},b} \equiv P_{0,b} + k_b\,x_{{\rm sat},b}$ enforces continuity at the saturation knee $x_{{\rm sat},b}$. Each bin thus has three free parameters: the intercept $P_{0,b}$, the slope $k_b$, and the knee position $x_{{\rm sat},b}$. We fitted Equation~\eqref{eq:probcal} to the measured hit-rates (bins with at least 100 pixels) by weighted least squares, using the binomial hit-rate uncertainties as weights. The best-fit parameters are listed in Table~\ref{tab:probfit} and shown as solid curves in Figure~\ref{fig:prob_calib}. The intercept decreases and the knee moves to larger $x$ with increasing redshift, while the saturation probability remains $P_{\rm sat}\approx0.68$--$0.76$ across all bins. The near-zero intercept has a direct physical meaning: a pixel with $x\simeq0$ carries essentially no stacked-filament signal, so its probability reduces to the base coincidence rate---the fraction of the projected map area that falls within the matching tolerance of a genuine 3D filament. This areal covering fraction is set by the degree of development of the cosmic web. At low redshift the filamentary network is more evolved, with thicker and more space-filling filaments, so even a signal-free pixel has a $\sim40\%$ chance of tracing a real filament; towards high redshift the web is more tenuous and sparsely sampled, and this baseline falls to $\sim20\%$. The declining intercept therefore encodes the hierarchical growth of LSS with cosmic time, rather than any artifact of the mapping.

\begin{table}[htbp]
    \centering
    \caption{Best-fit parameters of the piecewise linear$+$saturation probability mapping.}
    {\renewcommand{\arraystretch}{1.3}
    \begin{tabular}{ccccc}
    \hline
       Redshift bin & $P_0$ & $k$ & $x_{\rm sat}$ & $P_{\rm sat}$ \\
       \hline
        $0.50 < z < 0.82$ & 0.411 & 0.0520 & 6.08 & 0.727 \\
        $0.82 < z < 1.21$ & 0.399 & 0.0425 & 7.01 & 0.697 \\
        $1.21 < z < 1.91$ & 0.329 & 0.0507 & 8.48 & 0.760 \\
        $1.91 < z < 3.52$ & 0.230 & 0.0485 & 9.25 & 0.678 \\
       \hline
    \end{tabular}}
    \tablefoot{Calibrated against $\mathcal{F}_{\rm 3D,ref}$ ($n_{\sigma,\rm ref}=2.5$) in the pixel-value space $x\equiv\ln(1+w_{\rm fil})$ (Equation~\ref{eq:probcal}). Columns list the intercept $P_0$, the slope $k$, the saturation knee $x_{\rm sat}$, and the plateau probability $P_{\rm sat}=P_0+k\,x_{\rm sat}$.}
    \label{tab:probfit}
\end{table}

Applying Equation~\eqref{eq:probcal} pixel-by-pixel to the observed $w_{\rm fil}$ maps yields a calibrated filament-probability map for each slice. We released these as 67 FITS files, \texttt{wfil\_prob\_slice\_\{s\}.fits} ($s=0$--$66$), one per tomographic slice. Each map has the same geometry as the corresponding $w_{\rm fil}$ map, stored as \texttt{float32}, with pixel values giving the calibrated filament probability $P\in[0,1]$. Because the pixel scale is fixed at $250$~ckpc while the comoving field of view grows with redshift, the map dimensions increase from $110\times99$ pixels at slice~0 to $377\times340$ pixels at slice~66. The FITS header records the redshift bin (\texttt{Z\_BIN}, \texttt{Z\_RANGE}), the calibration space (\texttt{CALSPACE}$=\ln(1+w_{\rm fil})$), and the per-bin mapping parameters (\texttt{PCAL\_P0}, \texttt{PCAL\_K}, \texttt{PCALXSAT}, \texttt{PCAL\_PST}), so that the mapping applied to each map is fully reproducible from its header.

\section{Persistence distribution comparison}\label{app:persis_compare}

\begin{figure}
    \centering
    \includegraphics[width=\linewidth]{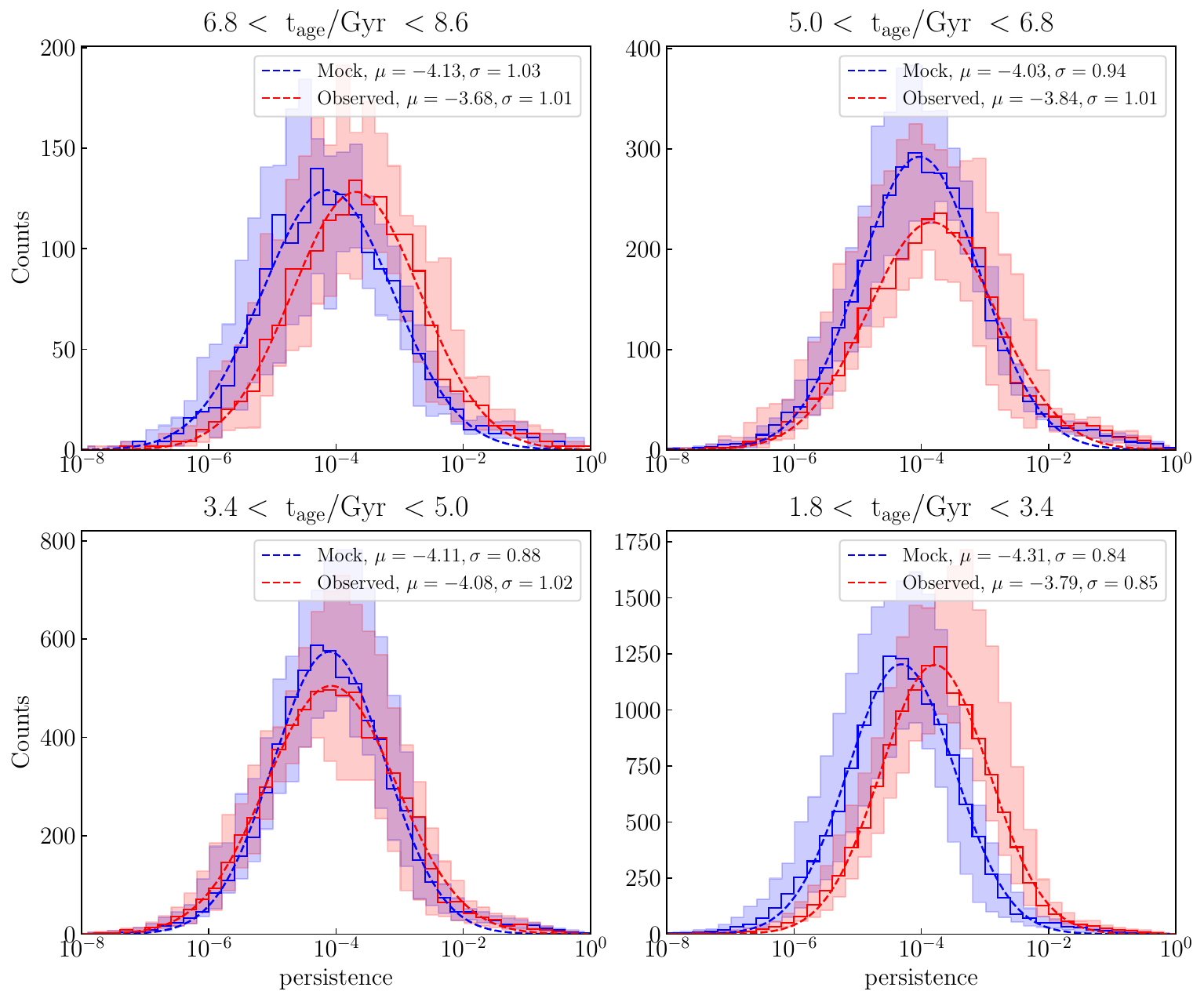}
    \caption{Distribution of persistence values for all persistence pairs connecting maxima and saddle points, derived from normalized $w_{\rm fil}$ maps, for the observed (red) and mock (blue) samples, separated into four redshift bins. The mean ($\mu$) and standard deviation ($\sigma$) of each distribution are reported in the legend. The shaded region indicates the range between the 16th and 84th percentiles in each bin.}
    \label{fig:persis_hist}
\end{figure}

Before determining the optimal $n_{\sigma,\rm pred}$ for the COSMOS-Web catalog, we verified that the mock and observed $w_{\rm fil}$ maps had comparable topological significance levels. Figure~\ref{fig:persis_hist} presents the persistence distributions for all maxima--saddle point pairs in both the mock and observed catalogs after normalizing the $w_{\rm fil}$ maps, divided into four redshift bins spanning an equal cosmic age of $\sim$1.8 Gyr: $0.50 < z < 0.82$, $0.82 < z < 1.21$, $1.21 < z < 1.91$, and $1.91 < z < 3.52$. All distributions are well described by Gaussian functions and, crucially, the mock and observed distributions are indistinguishable. This confirms that persistence thresholds calibrated on the mock catalog can be reliably applied to the observational data.

\section{Mock light cone construction details}\label{app:lightcone}

The mock light cone is generated following the procedure described in \cite{Bernyk2016}. Instead of simply filling the comoving space from redshift 0.5 to 3.5 with replicated cut of the simulation box, to avoid the periodicity caused by replicated large structures, we generated random cuts of the box by applying rotating, mirroring, and translation to the whole simulation box. For the detailed information about the transformation, see Chapter 4.2.2 in \cite{Bernyk2016}. Assembling randomly selected sub-boxes may introduce structural inconsistencies at the boundaries. However, \cite{Bernyk2016} show that the two-point correlation function is preserved relative to the original simulation volume. Furthermore, we have verified that these boundary inconsistencies do not significantly alter the statistical properties or the optimal parameters used for filament extraction. The procedure for generating random sub-boxes is detailed below.

To construct the mock light cone, the extracted sub-boxes are required to be non-overlapping, and thus the construction is based solely on even-numbered slices. For the i-th slice (i = 0, 2, 4, ..., 66) with redshift boundaries $z_{\rm i,min}$ and $z_{\rm i,max}$ (0.5 $\leq$ $z_{\rm i,min}$ < $z_{\rm i,max}$ < 3.5), assuming an observer located at the origin (0, 0, 0), we first determined the comoving Cartesian coordinate of the slice center as ($D_{\rm c,i}$, 0, 0), where $D_{\rm c,i}$ represents the comoving distance at $z$ = ($z_{\rm i,min}$ + $z_{\rm i,max}$) / 2. Subsequently, we identified the corresponding j-th simulation box (i.e. snapshot) positioned at ($D_{\rm c,sim,j}$ + $L_{\rm box}$ / 2, 0, 0), where $D_{\rm c,sim,j}$ denotes the comoving distance at $z$ = $z_{\rm sim,j}$ and $L_{\rm box}$ is the side length of the simulation box, such that the distance error $\Delta D_c$ = |$D_{\rm c,i}$ - ($D_{\rm c,sim,j}$ + $L_{\rm box}$ / 2)| is minimized. The correspondence between i and j is shown in Table~\ref{tab:info}.

We asserted the width and height of the field of view in COSMOS-Web DR1 to be $\rm \Delta R.A.$ = 0.6988$^\circ$ and $\rm \Delta Decl.$ = 0.7755$^\circ$. R.A. and Decl. of dark matter particles and subhalos in the mock light cone are defined as follows \citep{Ishikawa2024}:

\begin{equation}
    \mathrm{R.A._{sim} = arctan}(\frac{y_{ob}}{x_{ob}}),
\end{equation}

\begin{equation}
    \mathrm{Decl._{sim} = arcsin}(\frac{z_{ob}}{\mathbf{|r_{ob}|}}),
\end{equation}

where $\mathbf{r_{ob}}$ = ($x_{ob},y_{ob},z_{ob}$) is the position of a dark matter particle or a subhalo within the Cartesian coordinate with the origin at the observer. This position can be converted from the intrinsic Cartesian coordinates in the simulation box $\mathbf{r_0}$ = ($x_0,y_0,z_0$) by:

\begin{equation}
    x_{ob}=x_0+D_{\rm c,sim,j}+\frac{D_s}{2}[\mathbb{I}(A)-\mathbb{I}(B)],
\end{equation}

where $A$ = ${D_{\rm c,i}+D_s/2>D_{\rm c,sim,j}+L_{\rm box}}$, $B$ = ${D_{\rm c,i}-D_s/2\leq D_{\rm c,sim,j}}$, and $D_s$ = $1.5\times 10^5$ ckpc represents the slice thickness. $\mathbb{I}$ is the indicator function, $\mathbb{I}(A)$ = 1 if A is true, else $\mathbb{I}(A)$ = 0,

\begin{equation}
    y_{ob}=y_0-\frac{L_{\rm box}}{2},
\end{equation}

\begin{equation}
    z_{ob}=z_0-\frac{L_{\rm box}}{2}.
\end{equation}

Subsequently, we selected particles and subhalos inside sub-boxes based on the condition

\begin{equation}
    \left\{\begin{array}{rcl}
        D_{\rm c,i}-D_s/2< & \mathbf{|r_{ob}|}&<D_{\rm c,i}+D_s/2, \\
        -\rm{\Delta R.A.}/2< & \rm{R.A._{sim}}&<\rm{\Delta R.A.}/2, \\
        -\rm{\Delta Decl.}/2< & \rm{Decl._{sim}}&<\rm{\Delta Decl.}/2.
    \end{array}\right.
\end{equation}

We then applied this selection procedure to all 34 even-numbered slices and ultimately constructed the mock light cone, which matches the observational light cone in comoving size. The light cone is generated at each coarse time step of the simulation; however, structures evolve during time steps, which leads to unavoidable discontinuities at the boundaries between successive light cone slices. As reported by \cite{Pichon2010}, this discontinuity primarily arises from the thermal motion of particles within the largest cluster, with amplitudes below 200 ckpc (see Figure D1 in \cite{Pichon2010}). Given that the resolution of the $w_{\rm fil}$ maps is set to 250 ckpc, this discontinuity is not expected to significantly affect the statistical results of this study. Figure~\ref{fig:heatmap_dm} shows the front view and the side view of the first light cone slice.
\end{appendices}

\end{document}